\documentclass[floatfix,groupedaddress,superscriptaddress,aps,showpacs,amsmath,amssymb,floatfix, prl, longbibliography, twocolumn,a4]{revtex4-1}

\usepackage{graphicx,float}
\usepackage{subfigure}
\usepackage{dcolumn}
\usepackage{bm}
\usepackage{mathrsfs}
\usepackage{txfonts}
\usepackage{CJK}
\usepackage[amssymb]{SIunits}
\usepackage{epsfig}
\usepackage{epstopdf}
\usepackage{lipsum}
\usepackage{color}
\usepackage{placeins}

\usepackage[colorlinks,
linkcolor=blue,
anchorcolor=blue,
citecolor=blue,
filecolor=blue,
menucolor=blue,
runcolor=blue,
urlcolor=blue,
frenchlinks=blue]{hyperref}

\newenvironment{SChinese}{%
\CJKfamily{gbsn}%
\CJKtilde
\CJKnospace}{}

\allowdisplaybreaks[2]

\begin{document}

\begin{CJK}{UTF8}{}
\begin{SChinese}

\title{Towards On-Demand Heralded Single-Photon Sources via Photon Blockade}

\author{Jiangshan Tang}  %
 \affiliation{College of Engineering and Applied Sciences, National Laboratory of Solid State Microstructures, and Collaborative Innovation Center of Advanced Microstructures, Nanjing University, Nanjing 210023, China}
  \affiliation{School of Physics, Nanjing University, Nanjing 210023, China}

 \author{Lei Tang}  %
 \affiliation{College of Engineering and Applied Sciences, National Laboratory of Solid State Microstructures, and Collaborative Innovation Center of Advanced Microstructures, Nanjing University, Nanjing 210023, China}

  \author{Haodong Wu}  %

  \affiliation{College of Engineering and Applied Sciences, National Laboratory of Solid State Microstructures, and Collaborative Innovation Center of Advanced Microstructures, Nanjing University, Nanjing 210023, China}

\author{Yang Wu}  %
 \affiliation{College of Engineering and Applied Sciences, National Laboratory of Solid State Microstructures, and Collaborative Innovation Center of Advanced Microstructures, Nanjing University, Nanjing 210023, China}
 \affiliation{School of Physics and Information Technology, Shaanxi Normal University, Xi'an 710062, China}

\author{Hui Sun}  %
 \email{physunh@snnu.edu.cn}
 \affiliation{School of Physics and Information Technology, Shaanxi Normal University, Xi'an 710062, China}

\author{Han Zhang}  %
\email{zhanghan@nju.edu.cn}
   \affiliation{College of Engineering and Applied Sciences, National Laboratory of Solid State Microstructures, and Collaborative Innovation Center of Advanced Microstructures, Nanjing University, Nanjing 210023, China}
  \affiliation{School of Physics, Nanjing University, Nanjing 210023, China}

\author{Tao Li}  %
 \affiliation{School of Science, Nanjing University of Science and Technology, Nanjing 210094, China}

\author{Yanqing Lu (陆延青)}  %
     \affiliation{College of Engineering and Applied Sciences, National Laboratory of Solid State Microstructures, and Collaborative Innovation Center of Advanced Microstructures, Nanjing University, Nanjing 210023, China}
  \affiliation{School of Physics, Nanjing University, Nanjing 210023, China}

\author{Min Xiao}  %
   \affiliation{College of Engineering and Applied Sciences, National Laboratory of Solid State Microstructures, and Collaborative Innovation Center of Advanced Microstructures, Nanjing University, Nanjing 210023, China}
  \affiliation{School of Physics, Nanjing University, Nanjing 210023, China}
 \affiliation{Department of Physics, University of Arkansas, Fayetteville, Arkansas 72701, USA}

\author{Keyu Xia (夏可宇)}  %
 \email{keyu.xia@nju.edu.cn}
    \affiliation{College of Engineering and Applied Sciences, National Laboratory of Solid State Microstructures, and Collaborative Innovation Center of Advanced Microstructures, Nanjing University, Nanjing 210023, China}
  \affiliation{School of Physics, Nanjing University, Nanjing 210023, China}


\begin{abstract}
Spontaneous parametric down-conversion (SPDC) in a laser pumped optical nonlinear medium can produce heralded single photons with a high purity but a very low yield. Improving the yield by increasing the pump power in SPDC inevitably reduces the purity due to excitation of multi-photon events.  We propose a scheme to overcome this purity-yield trade-off by suppressing multi-photon events in a cavity-enhanced SPDC via the photon blockade effect.
By introducing a strong photon-photon interaction into the intracavity medium and increasing the pump power, we can improve the available single-photon yield to larger than $90\%$, while maintaining a high purity of $99\%$, towards on-demand generation of single photons through the SPDC process. Our quasi-on-demand SPDC sources may boost single-photon-based quantum information technology.
\end{abstract}

\maketitle

\end{SChinese}
\end{CJK}

Single photons are at the heart of photon-based quantum information processing \cite{RevModPhys.79.135,nnano.2017.218,RevModPhys.89.035002} such as quantum communication \cite{RevModPhys.84.777}, quantum simulations \cite{RevModPhys.86.153} and linear optical quantum computing \cite{science.abe8770}. The quality of single-photon sources directly determines the development and performance of photon-based quantum information technology \cite{RevModPhys.87.347,PhysRevLett.121.250505,PhysRevLett.121.203602,science.aba9779}.

Various methods have been studied for generating single photons, including photon blockade of a weak coherent field \cite{PhysRevLett.79.1467}, nonlinear wave mixing \cite{OE.14.012388, PhysRevA.83.054302,nphoton.2013.339, lsa.2017.100}, cavity quantum electrodynamics (CQED) \cite{nature06126, nphys2612,PhysRevLett.116.020401,nphoton.2016.23,nphoton.2019.10.1038}.
Among these methods, CQED systems using quantum dots (QDs) have achieved great success \cite{PhysRevLett.116.020401,nphoton.2016.23,nnano.2017.218,nphoton.2019.10.1038}, and promise a kind of high-performance single-photon source.
Some advanced techniques such as the dichromatic excitation \cite{nphysics.10.1038.941} and asymmetrical cavities are reported to overcome the challenging issues of polarization filtering \cite{nphoton.2019.10.1038}, spectral overlap with the exciting laser \cite{PhysRevLett.126.047403}.

Another widely used and important approach of generating single photons is  the heralded single-photon source (HSPS) based on spontaneous parametric down-conversion (SPDC) in an optical nonlinear medium. It provides another commonly used approach for many important room-temperature applications in quantum information technologies, in particular, requiring a narrow-band photon sources  \cite{PhysRevLett.83.2556,nphoton.2011.213,lsa.2016.249, PhysRevLett.125.263602}.
HSPSs are almost perfect in all aspects such as indistinguishability \cite{nphoton.2016.23}, high purity \cite{nnano.2017.218}, flexibility \cite{lsa.2016.249}, and scalability \cite{ncomms3582}, except for the only drawback of the small photon yield, which is typically few parts per thousand for guaranteeing a high purity \cite{nnano.2017.218}.
Cavity-enhanced narrow-band SPDC can reduce the single-photon bandwidth to fit atomic systems and quantum memory \cite{PhysRevLett.83.2556, nphoton.2011.213}, which normally operates in megahertz to gigahertz and at room temperature \cite{nphoton.2010.30}.

There is a trade-off between the purity and the yield per pump pulse, namely the purity-yield limitation, in HSPSs \cite{nnano.2017.218,PhysRevA.85.023829,njp.043030}. To produce heralded single photons with a high single-photon purity, the pump laser in the conventional HSPS must be weak enough that the probability of multi-photon emissions is negligibly small. Inevitably, a weak pumping also leads to a small generation probability of single photons, i.e. low single-photon yield. This inherent constraint between high single-photon purity and large photon yield is the fundamental limitation of applications of HSPSs \cite{RevModPhys.84.777,JMO.2010.546894}.
It is worth noting that multiplexed HSPS can improve the yield of signal photons without considerably reducing the purity by detecting a single idler photon from many individual HSPSs, different spectral components or a series of time bin of many SPDC processes \cite{PhysRevA.66.053805,PhysRevA.66.042303,PhysRevLett.119.083601,sciadv.aaw8586}.  However, overcoming the purity-yield trade-off still remains a challenge. A principle breakthrough of solving this inherent limitation is the key to boost applications of HSPSs, even open up many unprecedented single-photon-based quantum information technologies.

In this paper, we propose a scheme to improve the single-photon generation efficiency, namely yield per pump pulse, of HSPS based on the cavity-enhanced SPDC but keep a high purity, therefore overcoming the purity-yield limitation. By inserting warm N-type atoms in a ring optical cavity, we create a strong photon-photon interaction (PPI) to suppress the multi-photon pair events in SPDC under a strong pumping, and thus improve the yield of HSPSs from conventional value of few parts per thousand to more than $90\%$, whereas the purity simultaneously remains high.

 \begin{figure}
  \centering
  \includegraphics[width=1.0\linewidth]{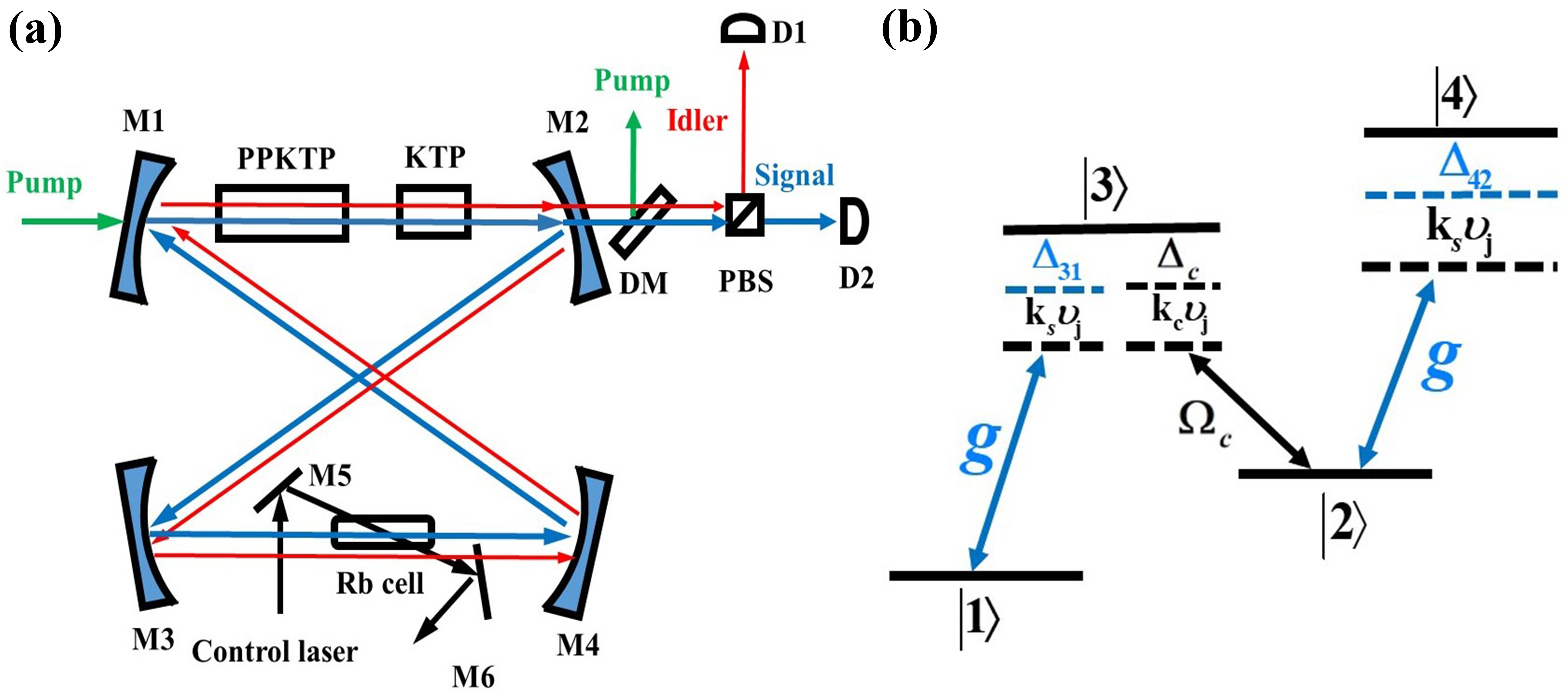} \\
\caption{(a) Schematic setup for overcoming the purity-yield trade-off in HSPS. A pump laser enters the ring cavity and drives a PPKTP crystal to produce idler-signal photon pairs. A KTP crystal enables dual-mode resonance for the idler and signal modes. The control laser is used to induce strong PPI in warm atoms for the signal mode. M1-M6: mirrors; DM: dichroic mirror; PBS: polarization beam splitter; D1 and D2: single-photon detectors. (b) Level diagram of N-type atoms. The signal photon (control field) drives the transitions $|1\rangle\leftrightarrow|3\rangle$ and $|2\rangle\leftrightarrow|4\rangle$ ( $|2\rangle\leftrightarrow|3\rangle$) with detuning $\Delta_{31}$ and $\Delta_{42}$ ($\Delta_{c}$), and corresponding coupling strength $g$ ($\Omega_c$).}
\label{fig:FIG1}
\end{figure}

Our system overcoming the purity-yield trade-off in cavity-enhanced HSPS via the photon blockade is depicted in Fig.~\ref{fig:FIG1}(a). A pump laser beam with frequency $\omega_p$ and effective pump power $\xi\sqrt{P}$ is input into the ring optical cavity formed with reflective mirrors M1-M4. It drives a type-II phase-matched periodically poled $\text{KTiOPO}_{4}$ (PPKTP) crystal to induce an optical SPDC process \cite{nphoton.2011.213}.  This SPDC process emits photons in pairs: the idler and signal modes $a_i$ and $a_s$ with frequency $\omega_i$ and $\omega_s$, respectively, doubly resonant with the cavity \cite{OE.15.007940,PhysRevLett.101.190501,PhysRevLett.108.210501}.
The counter-propagating cavity modes are decoupled from the SPDC due to the chirality in frequency conversion.
The idler mode is orthogonal in polarization to and different in frequency from the signal mode. It is off-resonance with atomic transitions and thus decouples from Rb atoms. The pump laser is separated by the DM from the generated photons. The idler photon outputting from M2 is reflected by the PBS to the idler output channel and monitored by a single-photon detector D1. The signal mode couples to Rb atoms and is subject to strong PPI induced by the control laser field \cite{PhysRevA.84.053820}. It is detected by another detector D2 after leaving the cavity through M2.

The N-type Rubidium (Rb) atoms embedded in the optical cavity at room temperature creates a large PPI for the signal mode under the driving of the control laser~\cite{PhysRevA.84.053820}. The energy levels of the N-type atom are denoted as $| l \rangle$ with eigenfrequency $\omega_l$ ($l = 1,2,3,4$),  as shown in Fig.~\ref{fig:FIG1}(b). The state $|3\rangle$ decays to the states $|1\rangle$ and $|2\rangle$ with rates $\gamma_{31}$ and $\gamma_{32}$, respectively. The state $|4\rangle$ decays at a rate $\gamma_{42}$. The dephasing rates of both ground states $|1\rangle$ and $|2\rangle$ are $\gamma_{21}$. For simplicity, we assume $\gamma_{31}=\gamma_{32}=\gamma_{42}=\gamma_0$ and $\gamma_{21}\ll \gamma_0$ \cite{PhysRevLett.117.203601,PhysRevLett.121.203602,PhysRevLett.125.243601}. The cavity mode couples to the transitions $|1\rangle\leftrightarrow|3\rangle$ and $|2\rangle\leftrightarrow|4\rangle$ with strength $\text{g}$. The control laser beam has an angular frequency $\omega_c$, and drives the transition $|2\rangle\leftrightarrow|3\rangle$ with Rabi frequency $\Omega_c$. In the rotating frame, defined by an unitary transformation $U_1=\text{exp}\{i\sum_{j=1}^N[\omega_1\sigma_{11}^{j}+(\omega_s+\omega_1-\omega_c)\sigma_{22}^{j}+(\omega_s+\omega_1)\sigma_{33}^{j}+(2\omega_s+\omega_1-\omega_c)\sigma_{44}^{j}]t+i(\omega_ia_i^\dag a_i+\omega_sa_s^\dag a_s)t\}$, the Hamiltonian of the system takes the form ($\hbar = 1$)
\begin{equation}
   \label{eq:EQ1}
   \begin{split}
  H=&\sum_{j=1}^N \sum_{l=2}^{4} \Delta_{ll}\sigma_{ll}^{j} +\xi\sqrt{P}(e^{-i\Delta_p t}a_i^\dag a_s^\dag+e^{i\Delta_p t}a_ia_s)\\
&+ \sum_{j=1}^N \left[ g(a_s^\dag\sigma_{13}^{j}+a_s^\dag\sigma_{24}^{j})+\Omega_c\sigma_{23}^{j}+h.c. \right]\;,
\end{split}
\end{equation}
with $\Delta_{22}= \Delta_c- \Delta_{31}$, $\Delta_{33}=- \Delta_{31}$, $\Delta_{44} =  \Delta_c - \Delta_{42} - \Delta_{31} $, and $\Delta_p = \omega_p -\omega_s -\omega_i$.
$\sigma_{mn}^{j}=|m\rangle\langle n|~(m,n=1,2,3,4)$ denotes the transition operator for the $j$th atom. $\Delta_{31}=\omega_s-\omega_{31}$ ( $\Delta_{42}=\omega_s-\omega_{42}$ ) is the detuning between the cavity mode and the transition of $|3\rangle\leftrightarrow|1\rangle$ ( $|4\rangle\leftrightarrow|2\rangle$ ), and $\Delta_c=\omega_c-\omega_{32}$ is the detuning between the control field and the $|3\rangle\leftrightarrow|2\rangle$ transition ($\omega_{mn}=\omega_m-\omega_n$). The first term in Eq.~(\ref{eq:EQ1}) is the free Hamiltonian of the system. The second term represents the pump field applied to the PPKTP crystal \cite{RevModPhys.83.33}. The second line represents interaction of atoms with the cavity mode and the control laser field. Only the signal mode in the cavity interact with atoms. At room temperature, the inevitable random thermal motion of the $j$th atom moving with velocity $v_j$ causes the ``microscopic'' Doppler shifts $k_s v_j$ and $k_c v_j$, corresponding to wave vectors $k_s$ and $k_c$ of the cavity and control fields, respectively. Here we have $|k_s| \approx |k_c|$ and approximately take both as $k$. The control  and the cavity fields propagate in the same direction that $k_sv_j \approx k_cv_j$ \cite{PhysRevLett.121.203602}. We consider the condition of $|g| \ll |\Omega_c|$, leading to the population of the atomic ground state $\rho_{11}\approx1$.  Then we can use the perturbation approach to solve the master equation and obtain the effective PPI coefficient in the presence of thermal motion as \cite{PhysRevA.101.053802,scully1997quantum,okamoto2006fundamentals,PhysRevA.84.053820,PhysRevLett.121.203602} (see Supplement 1)
\begin{equation}
   \label{eq:EQ2}
   \begin{split}
  \eta=&\int\frac{ig^4\gamma_{32}}{|\Omega_c|^4\gamma_{21}(\gamma_{31}+\gamma_{32})}(\frac{1}{F(v)}+\frac{1}{F^*(v)})[\frac{(2\gamma_{21}+\gamma_{32})}{\gamma_{32}}\\
  &(\frac{1}{F^*(v)}-\frac{1}{F(v)})+(\frac{1}{F_1(v)}-\frac{1}{{F_1}^*(v)})] N(v)dv\;,
  \end{split}
  \end{equation}
where $F=-1/(i\Delta_{22} +\gamma_{21}/2)-(i(\Delta_{31}+kv)+(\gamma_{31}+\gamma_{32})/2)/|\Omega_c|^2$,
$F_1=1/(-i(\Delta_{42}-\Delta_c)-(\gamma_{42}+\gamma_{31}+\gamma_{32})/2)-(i(\Delta_{42}+kv)+(\gamma_{21}+\gamma_{42})/2)/|\Omega_c|^2$.
The atomic velocity has the Maxwell-Boltzmann distribution that $N(v)=N_a e^{-v^2/u^2}/u\sqrt{\pi}$, where $u$ is the room-mean-square atomic velocity, and $N_a$ is the total atom number. At room temperature, $k u\approx300~\mega\hertz$ for Rb atoms~\cite{PhysRevLett.100.173602}. This Kerr nonlinearity causes the PPI. This perturbation approach provides an estimation of Kerr nonlinearity in good agreement with experimental observations \cite{PhysRevLett.87.073601,PhysRevLett.101.073602,PhysRevA.84.053820,Wang_2013,PhysRevLett.117.203601,PhysRevLett.125.243601,PhysRevResearch.2.033517}. Moreover, the Kerr nonlinearity can be enhanced by increasing the number of atoms or adjusting $\Omega_c$.

Applying an unitary transformation of $U_2=\text{exp}\{i(\omega_p-\omega_i-\omega_s)a_s^\dag a_st\}$ to the system,  we get the effective Hamiltonian
\begin{equation}
   \label{eq:EQ3}
 H_\text{eff}=(\Delta-\eta-\Delta_p)a_s^\dag a_s+\eta (a_s^\dag a_s)^{2}+\xi\sqrt{P}(a_i^\dag a_s^\dag+a_ia_s)\;,
  \end{equation}
where $\Delta=\int N(v)dv[(g^2/{\Omega_c}^2)(\gamma_{32}(\Delta_{31}-\Delta_c)/(\gamma_{21}(\gamma_{31}+\gamma_{32}))(1/F+1/F^*)+(\Delta_{31}+kv)/(\gamma_{31}+\gamma_{32}))(1/F+1/F^*)+i(1/F^*-1/F))]$.  The influence of the unwanted linear frequency shift $\Delta$ can be eliminated by selecting an appropriate pump frequency $\omega_p$ that $\Delta = \Delta_p$.
The PPI causes an energy shift $\eta (a_s^\dag a_s)^2$ to the cavity mode $a_s$. This energy shift depends on the cavity excitation $\langle a_s^\dag a_s\rangle$. Once the cavity mode is excited to include a on-resonance signal photon ($\Delta_p = \Delta$), exciting a second signal photon in the cavity requires additional $2\eta$ energy and is suppressed. This photon blockade effect can be used to suppress multi-photon events in the SPDC.

The master equation of the density matrix $\rho$ describing the dynamics of our system is given by $\dot{\rho}(t)=-i\left[H_\text{eff},\rho(t)\right]+ \mathcal{L} \{\kappa_i,a_i\}\rho + + \mathcal{L} \{\kappa_s,a_s\}\rho$,
where $\kappa_i=\kappa_{i,ex}+\kappa_{i,in}$ ($\kappa_s=\kappa_{s,ex}+\kappa_{s,in}$) is the decay rate of the idler (signal) mode. The subscripts ``ex" and ``in" indicate the external rate and the internal loss rate, respectively. We assume $\kappa_{i,ex}=\kappa_{s,ex}=\kappa_{ex}$ and $\kappa_{i,in}=\kappa_{s,in}=\kappa_{in}$. The Lindblad term is $\mathcal{L}\{\kappa_l, a_l\}\rho = \kappa_l\left[2a_l\rho(t)a_l^\dag-a_l^\dag a_l\rho(t)-\rho(t)a_l^\dag a_l\right]$  with $l \in \{i, s\}$. Using the density matrix, we can evaluate the photon states in the cavity. After the pump pulse is switched off, the photons escape from the cavity. Then, the outgoing photons are detected.

In contrast, the quantum trajectory simulation provides information of emitted photon wavefunctions~\cite{RevModPhys.70.101,nature05589,Com.Phys.Coms.183}. By averaging many trajectories, we can calculate the population of each photonic state in the outgoing wavefunctions by counting the collapse of the system state over a period $\kappa_s t=6$ (see Supplement 1).  This period allows the generated photons to completely escape from the cavity through mirror M2 and to be detected.

The escape efficiency of signal (idler) photons is evaluated as $\eta_{esc,s(i)}=\kappa_\text{ex}/(\kappa_\text{ex}+\kappa_\text{in})$ \cite{PhysRevLett.123.133602}. In a waveguide-based HSPS with complicate structure, the mode mismatching and the system internal loss result in a low escape efficiency \cite{Luo_2015}. In our cavity-based HSPS, the cavity mode profile is axial symmetric and can match the geometry of output mirror well. The cavity internal loss can be  small \cite{Luo_2015} and the escape efficiency approaches to unity \cite{OE.15.007940,OE.16.018145,OL.41.005341}. To focus on the key physics of interest, we consider the ideal case $\eta_{esc,s(i)}=1$ that the photons are completely extracted through mirror M2.

The purity of the HSPS can be evaluated through the equal time second-order autocorrelation function $g^{(2)}_s (0)$ of the signal mode, conditional on the detection of the idler photon \cite{nnano.2017.218}. In our setup [see Fig.~\ref{fig:FIG1}(a)], once an idler photon is detected by D1, the system state degrades to a reduced density matrix $\rho_s$. The autocorrelation function $g^{(2)}_s (0)$ is given by \cite{Fasel_2004,PhysRevLett.102.063603,PhysRevA.79.035801} (see Supplement 1)
\begin{equation}
   \label{eq:EQ5}
 g_{s}^{(2)}(0)=\frac{\langle a_s^\dag(t)a_s^\dag(t)a_s(t)a_s(t)\rangle_{i}}{\langle a_s^\dag(t)a_s(t)\rangle^2_{i}}=\frac{\sum_{n_s\geq1} n_s(n_s-1)\alpha_{n_s}}{(\sum_{n_s}n_s\alpha_{n_s})^2}\;,
  \end{equation}
 where $\langle\cdots\rangle_{i}$ is the average over the post-measurement state when an idler photon is measured. $\alpha_{n_s}=\sum_{n_i>0}P_{n_s,n_i}/\sum_{n_{s}^{'},n_{i}^{'}>0}P_{n_{s}^{'},n_{i}^{'}}$ is the $n$-signal-photon probability after detecting an idler photon. $P_{n_s,n_i}$ is the probability of the states $|n_s,n_i\rangle$ under the Fock states representation of the associating system of the idler and signal modes. The output signal mode is antibunching if $g_{s}^{(2)}(0)<1$.

 In conventional HSPSs, improving the photon yield by increasing the pump laser power can result in reduction of the single-photon purity.
 This inherent but critical issue fundamentally limits scaling up of HSPS-based applications~\cite{nnano.2017.218,sciadv.aaw8586}. By exploiting the photon blockade, we show a strong PPI can effectively suppress the generation of multi-photon pairs in the SPDC process and thus guarantees a high purity but allows us to greatly improve the production efficiency of heralded single photons by increasing the pump laser power.

Below is the understanding of the physical mechanism breaking the fundamental purity-yield limitation. When the signal mode is excited in the nonlinear cavity, excitation of high Fock states is blocked due to the photon blockade effect. In the SPDC process, the idler and signal photons are always generated in pairs. Therefore, if the signal mode can only be generated in the form of single photon, the idler mode is also subject to the same situation. As a result,  the generation of the high Fock-state photon pairs $|n_s, n_i\rangle$ ($n_s=n_i \geq 2$) is also blocked. On the other hand, the probability of generating single-photon pairs becomes larger because the pump laser power is stronger. In this way, we fundamentally circumvent the inherent contradiction between remaining high purity and improving yield of HSPSs.
Detection of the idler photon with a probability close to unity means quasi-on-demand generation of a single signal photon with high purity.

\begin{figure}
  \centering
  \includegraphics[width=1.0\linewidth]{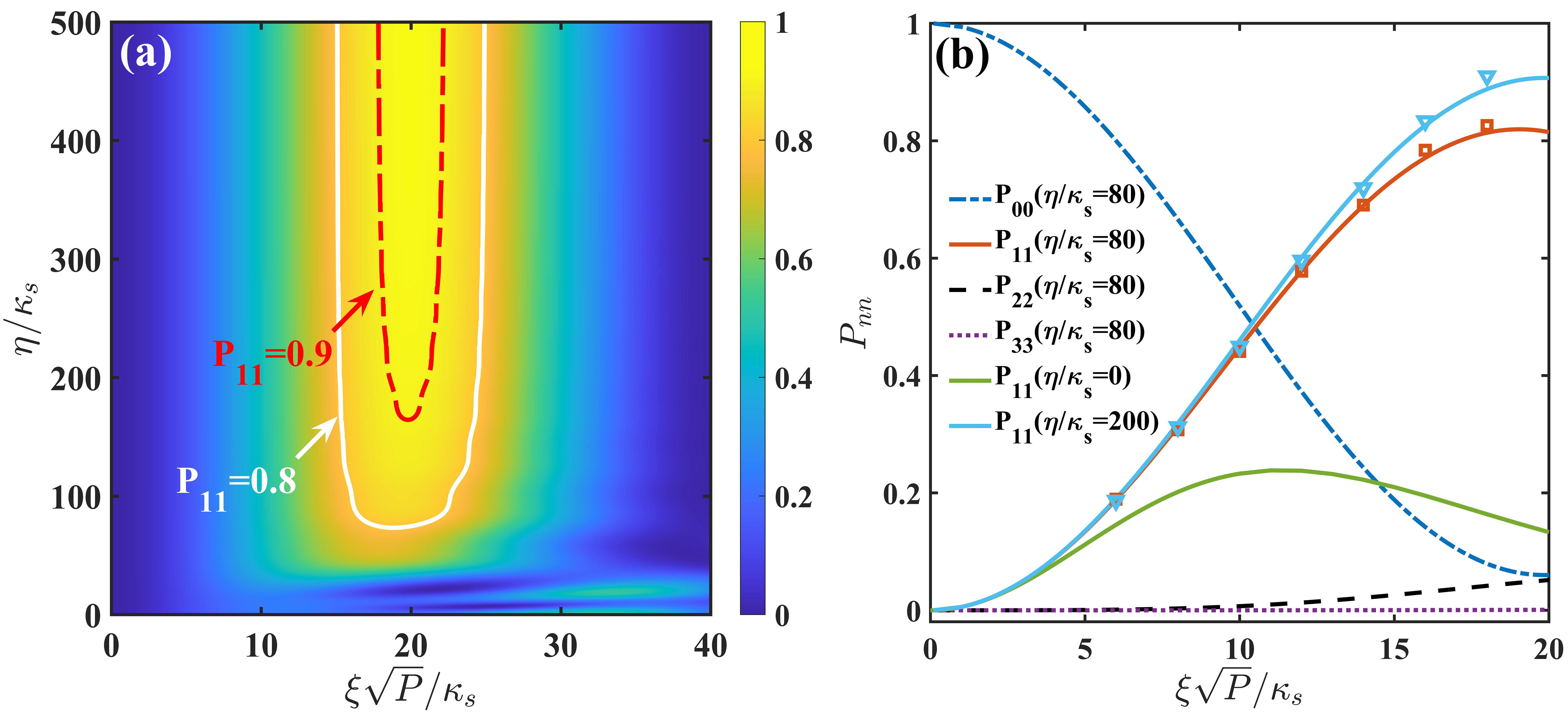} \\
  \caption{ (a) Population of Fock state $|1,1\rangle$, i.e., the probability of single-photon pair, $P_{11}$ versus the pump laser power $\xi\sqrt{P}$ and the effective nonlinear PPI strength $\eta$. (b) Population of paired Fock states $|n_s,n_i \rangle$ as a function of $\xi\sqrt{P}$ with different PPI strength. Other parameters are $\kappa_s=\kappa_i=1$.}
  \label{fig:FIG2}
\end{figure}

 \begin{figure}
  \centering
  \includegraphics[width=1.0\linewidth]{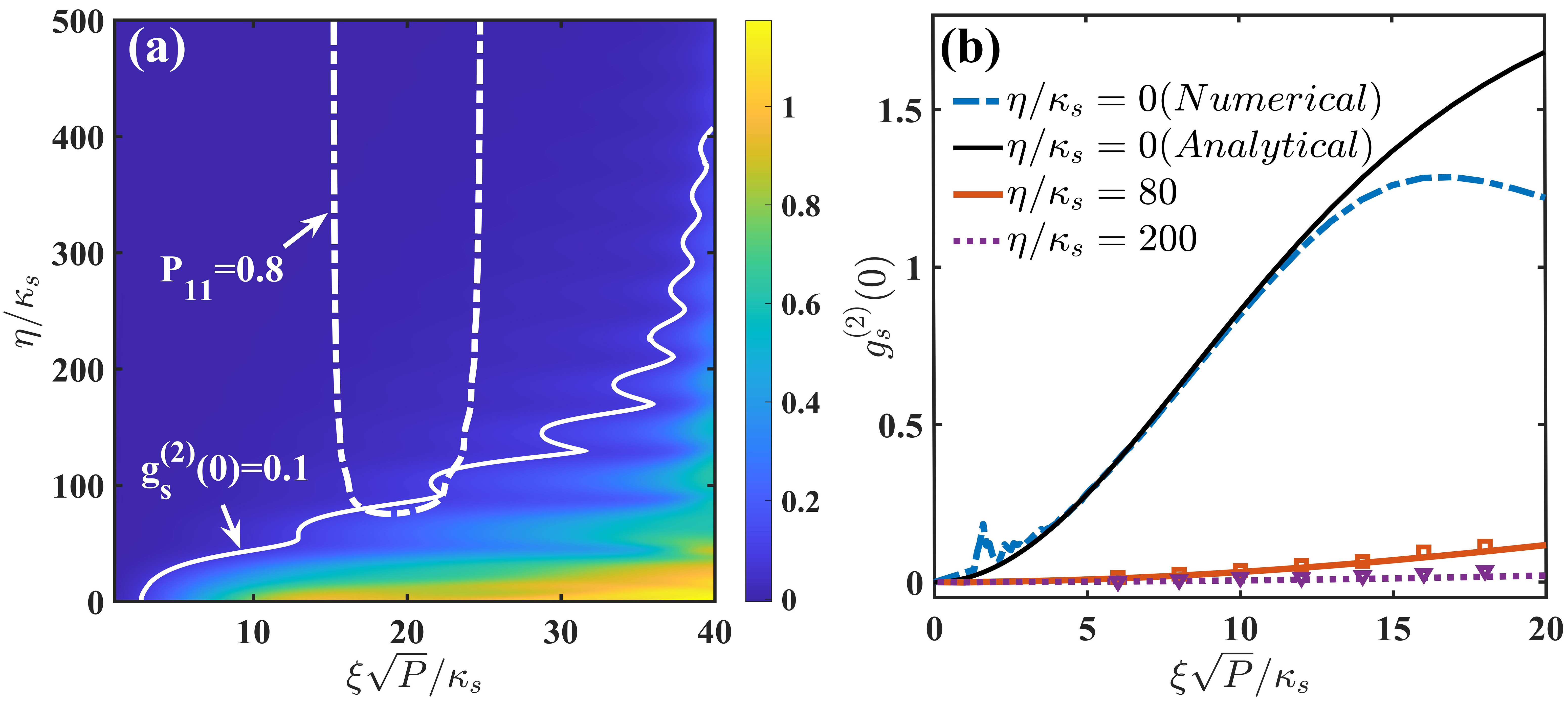} \\
  \caption{ (a) The equal-time second-order autocorrelation $g_s^{(2)}(0)$ versus the pump laser power $\xi\sqrt{P}$ and the PPI strength $\eta$. (b) The second-order autocorrelation $g_s^{(2)} (0)$ as a function of $\xi\sqrt{P}$ with different PPI strength $\eta$. Other parameters are the same as in Fig.~\ref{fig:FIG2}.}
  \label{fig:FIG3}
 \end{figure}

The evolution of the system can be obtained by solving the master equation. The probability of generating a single photon in each outgoing light pulse is a good measure of the yield $Y$ per pump pulse or brightness~\cite{nnano.2017.218}. If $P_{n}$ is the $n$-photon number probability, then $Y=P_1$ for $P_1\gg P_{n>1}$ and $Y=1$ for a deterministic single-photon source~\cite{nnano.2017.218}.  In our system, as signal photons and idler photons are created in pairs, then $Y=P_{11}$. For convenience, here the subscripts in $P_{n_s,n_i}$ represent signal and idler photon numbers, respectively. Note that the yield $Y$ is essentially different from the so-called single-photon heralding efficiency.  The later describes the probability of successfully heralding a signal photon when an idler photon is detected \cite{OL.41.005341,APL.5.0003601}. Even at a low yield of photon pair $|11\rangle$, high heralding efficiency can be obtained \cite{Fasel_2004,nphoton.2017.10.1038}.

Figure~\ref{fig:FIG2}(a) shows the population $P_{11}$ of Fock state $|1,1\rangle$ versus the pump laser power $\xi\sqrt{P}$ and the PPI strength $\eta$. The duration $\tau_p$ of the pump laser pulse is fixed at $\tau_p = \pi/40\kappa_s$. With the increase of $\eta$ to much larger than $\kappa_s$, the single-photon pair population $P_{11}$ can be significantly improved and peaks when $2\xi\sqrt{P} \tau_p = \pi$, corresponding to a $\pi$ pump pulse. For $\eta/\kappa_s>75$ and $15<\xi\sqrt{P}/\kappa_s<25$, indicated by the white solid curve, the yield can be higher than $80\%$. A larger PPI allows us to achieve $Y>90\%$, see the red dashed curve.
For a given PPI strength $\eta$, the population $P_{11}$ shows a cosine-like oscillation as the pump power increases, first increasing and then decreasing after reaching the maximum. Two reasons are behind this result: one is the excitation population oscillations as the pump pulse area $2\xi\sqrt{P}\tau_p$ increases, and the other is that the given PPI becomes not large enough to suppress the multi-photon pair generation excited by a strong pump field.

The available maximum yield is determined by the pump power $\xi\sqrt{P}$ and the PPI strength $\eta$, see Fig.~\ref{fig:FIG2}(b).
Enhancing the pump power can improve the single-photon yield, and increasing the PPI can suppress the multi-photon excitation.
Without nonlinearity-induced photon blockade, the probability of creating photon pairs $|n_s, n_i\rangle$ scales with pump intensity to the $n$th power as predicted theoretically \cite{nnano.2017.218}, see Fig.~\ref{fig:FIG2}(b). The maximum yield can only be $23.8\%$ at $\xi\sqrt{P}/\kappa_s \simeq 11$ at the expense of low purity ($\Pi \approx 4.3\%$) due to high excitation of multiphoton-pair states~\cite{OE.19.022698,PhysRevA.85.023829,njp.043030,PhysRevLett.119.083601,sciadv.aaw8586}.
With photon blockade, multi-photon events almost disappear, see the dashed black for $P_{22}$ and dotted purple curves for $P_{33}$ in Fig.~\ref{fig:FIG2}(b).
When $\eta/\kappa_s = 80$, we obtain the maximum probability $P_{11}$, i.e. the yield, of $81\%$ at $\xi\sqrt{P}/\kappa_s \simeq 19$. Meanwhile, the vacuum state population drops rapidly with the power. We can further improve the yield to $\text{0.9}$ for $\eta/\kappa_s=200$ and $\xi\sqrt{P}/\kappa_s=20$. When the yield is higher than $90\%$, it is reasonable to claim that the HSPS becomes quasi-on-demand \cite{nnanotechnol.10.1038.108}. By ``quasi-on-demand'', we mean each pump pulse can generate a single-photon pulse with a large probability \cite{Rew.Sci.Instrum.82.071101}.

The single-photon purity $\Pi$ is defined as the single-photon fraction of the total population of the heralding signal modes \cite{nphys4052,SanchezMunoz:18}.
 It can be quantified by the second-order autocorrelation function $g_{s}^{(2)}(0)$~\cite{nnano.2017.218,nphoton.2019.10.1038,nphys4052,APL.1.5020038,SanchezMunoz:18}.
For a high purity, we use the relation $\Pi \approx 1-g_{s}^{(2)}(0)$ \cite{PhysRevLett.116.020401} (see Supplement 1).

Figure~~\ref{fig:FIG3}(a) shows the correlation function $g_s^{(2)}(0)$ versus the PPI strength and the pump power. we can obtain a high purity $\Pi > 90\%$ when $\eta$ is large enough, see the white solid curve. In the overlap region marked by the curves showing $P_{11} >0.8$ and $g_s^{(2)}(0) <0.1$, we simultaneously obtain  $Y >80\%$ and $\Pi >90\%$.
A larger pump power leads to a decrease in the purity, because the probability of multi-photon events becomes higher as the pump power increases, see $P_{22}$ and $P_{33}$ in Fig.~\ref{fig:FIG2}(b). In contrast to conventional HSPSs, increasing the PPI strength $\eta$ in our scheme can suppress the excitation of multi-photon states, and thus can maintain the high purity.

In Fig.~\ref{fig:FIG3}(b), we show the equal time second-order autocorrelation function $g_s^{(2)}(0)$ as an indicator of the single-photon purity for different PPI strengths. When the PPI is absent or small,  $g_s^{(2)}(0)$ increases to larger than unity as the pump power increases, corresponding to the case of heralded single-photon generation under a strong pumping. Without the PPI, we derive $g_s^{(2)}(0)\approx2 \text{tanh}^2(\xi\sqrt{P}\tau_p)$~(see Supplement 1), in good agreement with the numerical result.
The purity-yield product is always bounded by $0.09$. The difference at high pump power is due to the truncation of Hilbert space in simulation (see Supplement 1). The purity increases rapidly as the PPI becomes larger. For example, when $\eta/\kappa_s=80$, the yield is already $81\%$ and the purity reaches $91\%$ ($g_s^{(2)}(0)=0.09$) at $\xi\sqrt{P}/\kappa_s=19$. If $\eta/\kappa_s=200$, the yield and the purity increase to $91\%$ and $99\%$, respectively.

\begin{figure}
  \centering
  \includegraphics[width=1.0\linewidth]{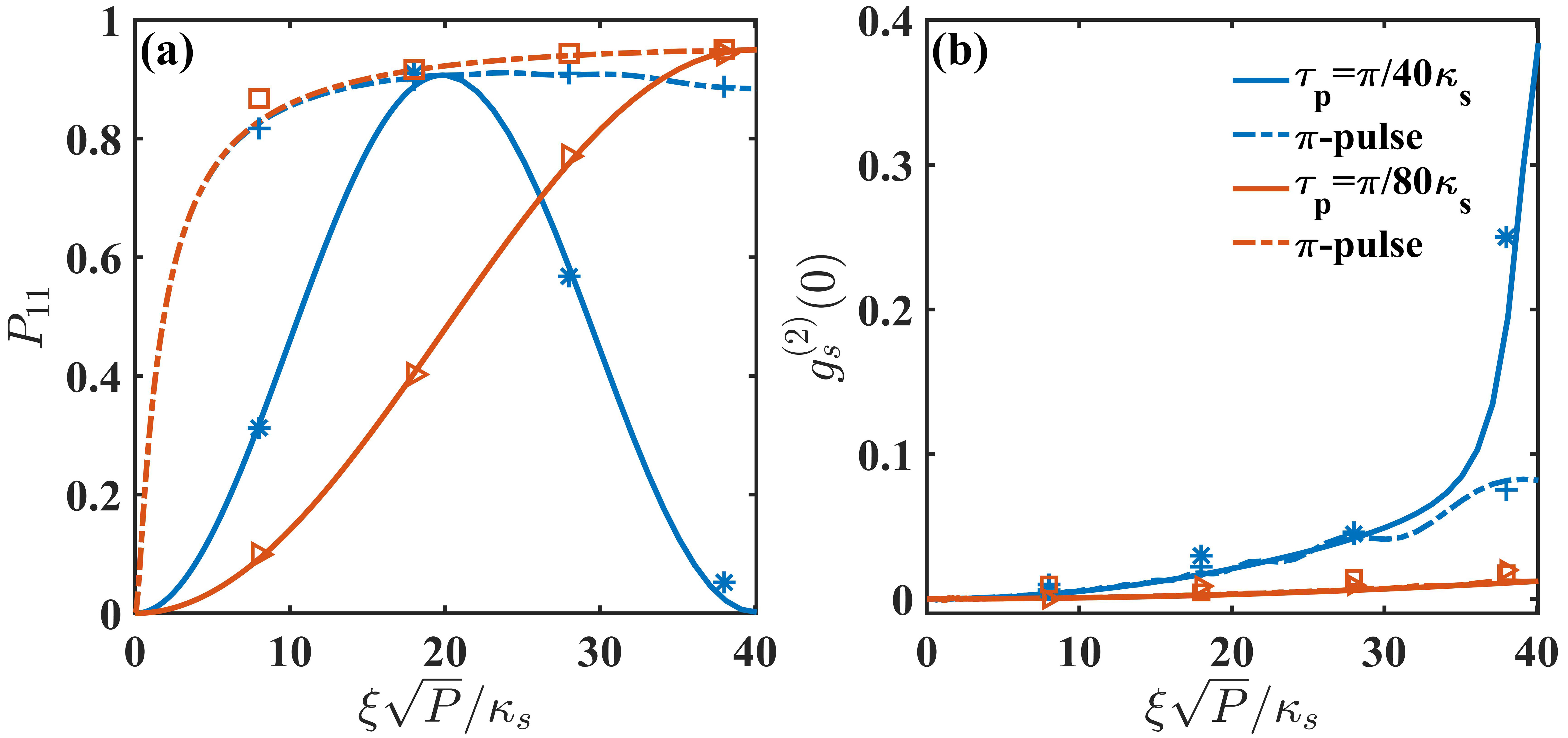} \\
  \caption{Population of Fock state $|1,1\rangle$ (a) and the second-order autocorrelation $g_s^{(2)}$ (b) versus the pump power $\xi\sqrt{P}$. Blue (red) solid curves indicate the case with a fixed pulse duration $\tau_p = \pi/40 \kappa_s$ and the PPI strength $\eta/\kappa_s=80$ ($\tau_p = \pi/80\kappa_s$ and $\eta/\kappa_s=500$). Blue (red) dash-dotted curves represent the results for $\pi$-pulse excitation, i.e. $2\xi\sqrt{P} \tau_p = \pi$, in the case of $\eta/\kappa_s=80$ ($\eta/\kappa_s=500$). Other parameters are $\kappa_s = \kappa_i =1$ and $\gamma_0=0.5\kappa_s$.}
  \label{fig:FIG4}
  \end{figure}

In our scheme, high Fock states of signal and idler modes are suppressed due to the photon blockade effect. To provide an apparent physical picture, we truncate the photonic basis to only include the vacuum and single-photon pair states in the cavity. In the effective two-state space, the state $|1,1\rangle$ can be completely populated from the vacuum state with a $\pi$ pump pulse~\cite{PhysRevLett.116.020401} (see Supplement 1).
In a real case including high Fock-state photon pairs, instead of fixing the pump pulse duration $\tau_p$ but changing the pump power $\xi\sqrt{P}$, we fix the pump pulse area $2\xi\sqrt{P} \tau_p =\pi$ to excite the single-photon pair $|1,1\rangle$ and investigate the influence of the pump power on the yield and purity of HSPS. As shown in Fig.~\ref{fig:FIG4}, the single-photon yield per pump pulse can increase to $80\%$ at $\xi\sqrt{P}/\kappa_s=6.6$ under $\pi$-pulse excitation, in comparison with a yield of $22.6\%$ in fix-duration excitation. At the same time, the purity remains nearly unchanged, larger than $99\%$ in both excitation cases.
For relative low pump power, the yield under $\pi$-pulse excitation is much higher than the case of fix-duration excitation. In contrast, the purities are very close. Thus, a $\pi$ pulse can excite the single-photon pair state more efficiently and a high purity is guaranteed by the photon blockade effect.

When the PPI is large enough, the nonlinear cavity can be modeled as a two-level system \cite{science.aaw1611}. Then, our HSPS is equivalent to a CQED-based single-photon source. Thus, the yield can approach unity in theory. To show this, we calculate the yield and the purity in a very large PPI regime. For example, when $\eta/\kappa_s=500$, we obtain $Y=95\%$ and $\Pi \approx 99\%$ ($g^{(0)}\approx 0.01$) at $\xi\sqrt{P}/\kappa_s=40$, see the red curves in Fig.~\ref{fig:FIG4}.

The generated photons are detected after escaping from the cavity. Our quantum trajectory simulations show that the quantum statistical characters of single-photon pairs outside the cavity is close to that of the single-photon pairs prepared in the cavity when the pump laser is switched off, see the simulation data markers at $\xi\sqrt{P}/\kappa_s=\{6, 8, 10, 12, 14, 16, 18\}$ in Figs.~\ref{fig:FIG2}(b) and ~\ref{fig:FIG3}(b), and at $\xi\sqrt{P}/\kappa_s=\{8,18,28,38\}$ in Fig.~\ref{fig:FIG4}.

Our scheme can be implemented by using a setup depicted in Fig.~\ref{fig:FIG1}(a). The high-quality ring optical cavity is formed with four reflective mirrors, M1-M4 ~\cite{PhysRevLett.74.666,OL.42.000271,nphoton.2018.1038}.
The mirror M$\text{2}$ has a relatively low reflectivity ($87.5\%$) as the output port \cite{OE.15.007940}. Other three mirrors have high reflectivity of $99.99\%$~\cite{OL.35.002293}.  End faces of the PPKTP and KTP crystals and the Rb atomic bubble are coated with $99.9\%$ anti-reflection layers for both the idler and signal modes. We use the ring cavity with a one-round length of $\text{0.8}\meter$. The internal losses of the cavity is calculated to be $\kappa_\text{in}=2\pi\times 0.37~\mega\hertz$ and the external loss rate is $\kappa_\text{ex}=2\pi\times8~\mega\hertz$. The escape efficiency can reach $\eta_{\text{esc},s(i)} = \kappa_\text{in}/(\kappa_\text{in} + \kappa_\text{ex})=95.6\%$.
An ensemble of warm Rb atoms is used to create a strong PPI for the signal mode $a_s$, with $\gamma_0 = 2\pi\times6~\mega\hertz$ \cite{PhysRevLett.121.203602}. Four levels related to the D1 line of $^{87}$Rb atom are used for the N-type configuration with $|1\rangle=|5^2S_{1/2},F=1,m_F=-1\rangle$, $|2\rangle=|5^2S_{1/2},F=2,m_F=0\rangle$, $|3\rangle=|5^2P_{1/2},F^{'}=2,m_{F}^{'}=-1\rangle$ and $|4\rangle=|5^2P_{1/2},F^{'}=2,m_{F}^{'}=0\rangle$. The signal mode and control laser drive the corresponding transitions  in Fig.~\ref{fig:FIG1}(b) with the detunings $\Delta_{31}=18\kappa_s\approx904~\mega\hertz$, $\Delta_{c}=0$ and $\Delta_{42}=8.5\Delta_{31}$. We can obtain the PPI strength of $\eta/\kappa_s = 200$ when $\Omega_c=30\gamma_0$ ($\text{0.9}\milli\watt$ \cite{PhysRevA.87.033835}) and $N_a (g/\kappa_s)^4 \sim 2.3 \times 10^3$. With $\pi$-excitation, we obtain the yield of $Y= 0.91$, the purity of $\Pi = 99\%$ and the escape efficiency of $\eta_{\text{esc},s}\eta_{\text{esc},i} = 91.4\%$, yielding a generation rate of $1$~\mega\hertz, when the total period of single-photon pulses takes $6\kappa_s^{-1}$. Alternatively, we can also use the N-type configuration in \cite{PhysRevA.84.053820} to achieve a strong PPI (see Supplement 1).

HSPSs and SPDC process have been realized in on-chip microring resonators \cite{lsa.2016.249,PhysRevLett.125.263602}. Atomic cladding photonic circuits have also been reported recently \cite{ncomms2554,APL.1.4927172,PhysRevX.8.021032}. Therefore, our scheme has the potential to be integrated on a chip if the microring resonator is cladded by Rb atoms. In such on-chip realization, the PPI can be significantly increased because the atom-light interaction in a microring resonator is much stronger. At same time, the outcoupling rate $\kappa_{ex}$ can be much larger than $\kappa_{in}$ if a bend coupling waveguide is applied \cite{PhysRevLett.125.263602}. Thus, we can expect better performance of our HSPS.

We have presented a method to solve the challenging problem of simultaneously achieving high purity and high yield in HSPSs. By exploiting the photon blockade to the signal mode in the cavity-enhanced HSPS, we have improved the single-photon yield of HSPS by two orders and keep the purity high. Considering the perfect characters of HSPSs like indistinguishability, bandwidth and flexibility, our quasi-on-demand HSPS may open a door for scalable photon-based quantum information processing.

 This work was supported by the National Key R\&D Program of China (Grants No. 2019YFA0308700, No. 2017YFA0303703, No. 2017YFA0303701), the National Natural Science Foundation of China (Grant Nos. 11874212, 11890704, 61671279,11574145,11690031), the Fundamental Research Funds for the Central Universities (021314380095) and the Jiangsu Innovation Plan. We thank the High Performance Computing Center of Nanjing University for doing the numerical calculations on its blade cluster system. We also thank Professor Xilin Wang for his helpful discussion.


\begin{thebibliography}{77}%
\makeatletter
\providecommand \@ifxundefined [1]{%
 \@ifx{#1\undefined}
}%
\providecommand \@ifnum [1]{%
 \ifnum #1\expandafter \@firstoftwo
 \else \expandafter \@secondoftwo
 \fi
}%
\providecommand \@ifx [1]{%
 \ifx #1\expandafter \@firstoftwo
 \else \expandafter \@secondoftwo
 \fi
}%
\providecommand \natexlab [1]{#1}%
\providecommand \enquote  [1]{``#1''}%
\providecommand \bibnamefont  [1]{#1}%
\providecommand \bibfnamefont [1]{#1}%
\providecommand \citenamefont [1]{#1}%
\providecommand \href@noop [0]{\@secondoftwo}%
\providecommand \href [0]{\begingroup \@sanitize@url \@href}%
\providecommand \@href[1]{\@@startlink{#1}\@@href}%
\providecommand \@@href[1]{\endgroup#1\@@endlink}%
\providecommand \@sanitize@url [0]{\catcode `\\12\catcode `\$12\catcode
  `\&12\catcode `\#12\catcode `\^12\catcode `\_12\catcode `\%12\relax}%
\providecommand \@@startlink[1]{}%
\providecommand \@@endlink[0]{}%
\providecommand \url  [0]{\begingroup\@sanitize@url \@url }%
\providecommand \@url [1]{\endgroup\@href {#1}{\urlprefix }}%
\providecommand \urlprefix  [0]{URL }%
\providecommand \Eprint [0]{\href }%
\providecommand \doibase [0]{http://dx.doi.org/}%
\providecommand \selectlanguage [0]{\@gobble}%
\providecommand \bibinfo  [0]{\@secondoftwo}%
\providecommand \bibfield  [0]{\@secondoftwo}%
\providecommand \translation [1]{[#1]}%
\providecommand \BibitemOpen [0]{}%
\providecommand \bibitemStop [0]{}%
\providecommand \bibitemNoStop [0]{.\EOS\space}%
\providecommand \EOS [0]{\spacefactor3000\relax}%
\providecommand \BibitemShut  [1]{\csname bibitem#1\endcsname}%
\let\auto@bib@innerbib\@empty
\bibitem [{\citenamefont {Kok}\ \emph {et~al.}(2007)\citenamefont {Kok},
  \citenamefont {Munro}, \citenamefont {Nemoto}, \citenamefont {Ralph},
  \citenamefont {Dowling},\ and\ \citenamefont {Milburn}}]{RevModPhys.79.135}%
  \BibitemOpen
  \bibfield  {author} {\bibinfo {author} {\bibfnamefont {P.}~\bibnamefont
  {Kok}}, \bibinfo {author} {\bibfnamefont {W.~J.}\ \bibnamefont {Munro}},
  \bibinfo {author} {\bibfnamefont {K.}~\bibnamefont {Nemoto}}, \bibinfo
  {author} {\bibfnamefont {T.~C.}\ \bibnamefont {Ralph}}, \bibinfo {author}
  {\bibfnamefont {J.~P.}\ \bibnamefont {Dowling}}, \ and\ \bibinfo {author}
  {\bibfnamefont {G.~J.}\ \bibnamefont {Milburn}},\ }\href {\doibase
  10.1103/RevModPhys.79.135} {\bibfield  {journal} {\bibinfo  {journal} {Rev.
  Mod. Phys.}\ }\textbf {\bibinfo {volume} {79}},\ \bibinfo {pages} {135}
  (\bibinfo {year} {2007})}\BibitemShut {NoStop}%
\bibitem [{\citenamefont {Senellart}\ \emph {et~al.}(2017)\citenamefont
  {Senellart}, \citenamefont {Solomon},\ and\ \citenamefont
  {White}}]{nnano.2017.218}%
  \BibitemOpen
  \bibfield  {author} {\bibinfo {author} {\bibfnamefont {P.}~\bibnamefont
  {Senellart}}, \bibinfo {author} {\bibfnamefont {G.}~\bibnamefont {Solomon}},
  \ and\ \bibinfo {author} {\bibfnamefont {A.}~\bibnamefont {White}},\ }\href
  {\doibase 10.1038/nnano.2017.218} {\bibfield  {journal} {\bibinfo  {journal}
  {Nat. Nanotechnol.}\ }\textbf {\bibinfo {volume} {12}},\ \bibinfo {pages}
  {1026} (\bibinfo {year} {2017})}\BibitemShut {NoStop}%
\bibitem [{\citenamefont {Degen}\ \emph {et~al.}(2017)\citenamefont {Degen},
  \citenamefont {Reinhard},\ and\ \citenamefont
  {Cappellaro}}]{RevModPhys.89.035002}%
  \BibitemOpen
  \bibfield  {author} {\bibinfo {author} {\bibfnamefont {C.~L.}\ \bibnamefont
  {Degen}}, \bibinfo {author} {\bibfnamefont {F.}~\bibnamefont {Reinhard}}, \
  and\ \bibinfo {author} {\bibfnamefont {P.}~\bibnamefont {Cappellaro}},\
  }\href {\doibase 10.1103/RevModPhys.89.035002} {\bibfield  {journal}
  {\bibinfo  {journal} {Rev. Mod. Phys.}\ }\textbf {\bibinfo {volume} {89}},\
  \bibinfo {pages} {035002} (\bibinfo {year} {2017})}\BibitemShut {NoStop}%
\bibitem [{\citenamefont {Pan}\ \emph {et~al.}(2012)\citenamefont {Pan},
  \citenamefont {Chen}, \citenamefont {Lu}, \citenamefont {Weinfurter},
  \citenamefont {Zeilinger},\ and\ \citenamefont {\ifmmode~\dot{Z}\else
  \.{Z}\fi{}ukowski}}]{RevModPhys.84.777}%
  \BibitemOpen
  \bibfield  {author} {\bibinfo {author} {\bibfnamefont {J.-W.}\ \bibnamefont
  {Pan}}, \bibinfo {author} {\bibfnamefont {Z.-B.}\ \bibnamefont {Chen}},
  \bibinfo {author} {\bibfnamefont {C.-Y.}\ \bibnamefont {Lu}}, \bibinfo
  {author} {\bibfnamefont {H.}~\bibnamefont {Weinfurter}}, \bibinfo {author}
  {\bibfnamefont {A.}~\bibnamefont {Zeilinger}}, \ and\ \bibinfo {author}
  {\bibfnamefont {M.}~\bibnamefont {\ifmmode~\dot{Z}\else \.{Z}\fi{}ukowski}},\
  }\href {\doibase 10.1103/RevModPhys.84.777} {\bibfield  {journal} {\bibinfo
  {journal} {Rev. Mod. Phys.}\ }\textbf {\bibinfo {volume} {84}},\ \bibinfo
  {pages} {777} (\bibinfo {year} {2012})}\BibitemShut {NoStop}%
\bibitem [{\citenamefont {Georgescu}\ \emph {et~al.}(2014)\citenamefont
  {Georgescu}, \citenamefont {Ashhab},\ and\ \citenamefont
  {Nori}}]{RevModPhys.86.153}%
  \BibitemOpen
  \bibfield  {author} {\bibinfo {author} {\bibfnamefont {I.~M.}\ \bibnamefont
  {Georgescu}}, \bibinfo {author} {\bibfnamefont {S.}~\bibnamefont {Ashhab}}, \
  and\ \bibinfo {author} {\bibfnamefont {F.}~\bibnamefont {Nori}},\ }\href
  {\doibase 10.1103/RevModPhys.86.153} {\bibfield  {journal} {\bibinfo
  {journal} {Rev. Mod. Phys.}\ }\textbf {\bibinfo {volume} {86}},\ \bibinfo
  {pages} {153} (\bibinfo {year} {2014})}\BibitemShut {NoStop}%
\bibitem [{\citenamefont {Zhong}\ \emph {et~al.}(2020)\citenamefont {Zhong},
  \citenamefont {Wang}, \citenamefont {Deng}, \citenamefont {Chen},
  \citenamefont {Peng}, \citenamefont {Luo}, \citenamefont {Qin}, \citenamefont
  {Wu}, \citenamefont {Ding}, \citenamefont {Hu}, \citenamefont {Hu},
  \citenamefont {Yang}, \citenamefont {Zhang}, \citenamefont {Li},
  \citenamefont {Li}, \citenamefont {Jiang}, \citenamefont {Gan}, \citenamefont
  {Yang}, \citenamefont {You}, \citenamefont {Wang}, \citenamefont {Li},
  \citenamefont {Liu}, \citenamefont {Lu},\ and\ \citenamefont
  {Pan}}]{science.abe8770}%
  \BibitemOpen
  \bibfield  {author} {\bibinfo {author} {\bibfnamefont {H.-S.}\ \bibnamefont
  {Zhong}}, \bibinfo {author} {\bibfnamefont {H.}~\bibnamefont {Wang}},
  \bibinfo {author} {\bibfnamefont {Y.-H.}\ \bibnamefont {Deng}}, \bibinfo
  {author} {\bibfnamefont {M.-C.}\ \bibnamefont {Chen}}, \bibinfo {author}
  {\bibfnamefont {L.-C.}\ \bibnamefont {Peng}}, \bibinfo {author}
  {\bibfnamefont {Y.-H.}\ \bibnamefont {Luo}}, \bibinfo {author} {\bibfnamefont
  {J.}~\bibnamefont {Qin}}, \bibinfo {author} {\bibfnamefont {D.}~\bibnamefont
  {Wu}}, \bibinfo {author} {\bibfnamefont {X.}~\bibnamefont {Ding}}, \bibinfo
  {author} {\bibfnamefont {Y.}~\bibnamefont {Hu}}, \bibinfo {author}
  {\bibfnamefont {P.}~\bibnamefont {Hu}}, \bibinfo {author} {\bibfnamefont
  {X.-Y.}\ \bibnamefont {Yang}}, \bibinfo {author} {\bibfnamefont {W.-J.}\
  \bibnamefont {Zhang}}, \bibinfo {author} {\bibfnamefont {H.}~\bibnamefont
  {Li}}, \bibinfo {author} {\bibfnamefont {Y.}~\bibnamefont {Li}}, \bibinfo
  {author} {\bibfnamefont {X.}~\bibnamefont {Jiang}}, \bibinfo {author}
  {\bibfnamefont {L.}~\bibnamefont {Gan}}, \bibinfo {author} {\bibfnamefont
  {G.}~\bibnamefont {Yang}}, \bibinfo {author} {\bibfnamefont {L.}~\bibnamefont
  {You}}, \bibinfo {author} {\bibfnamefont {Z.}~\bibnamefont {Wang}}, \bibinfo
  {author} {\bibfnamefont {L.}~\bibnamefont {Li}}, \bibinfo {author}
  {\bibfnamefont {N.-L.}\ \bibnamefont {Liu}}, \bibinfo {author} {\bibfnamefont
  {C.-Y.}\ \bibnamefont {Lu}}, \ and\ \bibinfo {author} {\bibfnamefont {J.-W.}\
  \bibnamefont {Pan}},\ }\href
  {https://science.sciencemag.org/content/370/6523/1460} {\bibfield  {journal}
  {\bibinfo  {journal} {Science}\ ,\ \bibinfo {pages} {eabe8770}} (\bibinfo
  {year} {2020})}\BibitemShut {NoStop}%
\bibitem [{\citenamefont {Lodahl}\ \emph {et~al.}(2015)\citenamefont {Lodahl},
  \citenamefont {Mahmoodian},\ and\ \citenamefont
  {Stobbe}}]{RevModPhys.87.347}%
  \BibitemOpen
  \bibfield  {author} {\bibinfo {author} {\bibfnamefont {P.}~\bibnamefont
  {Lodahl}}, \bibinfo {author} {\bibfnamefont {S.}~\bibnamefont {Mahmoodian}},
  \ and\ \bibinfo {author} {\bibfnamefont {S.}~\bibnamefont {Stobbe}},\ }\href
  {\doibase 10.1103/RevModPhys.87.347} {\bibfield  {journal} {\bibinfo
  {journal} {Rev. Mod. Phys.}\ }\textbf {\bibinfo {volume} {87}},\ \bibinfo
  {pages} {347} (\bibinfo {year} {2015})}\BibitemShut {NoStop}%
\bibitem [{\citenamefont {Zhong}\ \emph {et~al.}(2018)\citenamefont {Zhong},
  \citenamefont {Li}, \citenamefont {Li}, \citenamefont {Peng}, \citenamefont
  {Su}, \citenamefont {Hu}, \citenamefont {He}, \citenamefont {Ding},
  \citenamefont {Zhang}, \citenamefont {Li}, \citenamefont {Zhang},
  \citenamefont {Wang}, \citenamefont {You}, \citenamefont {Wang},
  \citenamefont {Jiang}, \citenamefont {Li}, \citenamefont {Chen},
  \citenamefont {Liu}, \citenamefont {Lu},\ and\ \citenamefont
  {Pan}}]{PhysRevLett.121.250505}%
  \BibitemOpen
  \bibfield  {author} {\bibinfo {author} {\bibfnamefont {H.-S.}\ \bibnamefont
  {Zhong}}, \bibinfo {author} {\bibfnamefont {Y.}~\bibnamefont {Li}}, \bibinfo
  {author} {\bibfnamefont {W.}~\bibnamefont {Li}}, \bibinfo {author}
  {\bibfnamefont {L.-C.}\ \bibnamefont {Peng}}, \bibinfo {author}
  {\bibfnamefont {Z.-E.}\ \bibnamefont {Su}}, \bibinfo {author} {\bibfnamefont
  {Y.}~\bibnamefont {Hu}}, \bibinfo {author} {\bibfnamefont {Y.-M.}\
  \bibnamefont {He}}, \bibinfo {author} {\bibfnamefont {X.}~\bibnamefont
  {Ding}}, \bibinfo {author} {\bibfnamefont {W.}~\bibnamefont {Zhang}},
  \bibinfo {author} {\bibfnamefont {H.}~\bibnamefont {Li}}, \bibinfo {author}
  {\bibfnamefont {L.}~\bibnamefont {Zhang}}, \bibinfo {author} {\bibfnamefont
  {Z.}~\bibnamefont {Wang}}, \bibinfo {author} {\bibfnamefont {L.}~\bibnamefont
  {You}}, \bibinfo {author} {\bibfnamefont {X.-L.}\ \bibnamefont {Wang}},
  \bibinfo {author} {\bibfnamefont {X.}~\bibnamefont {Jiang}}, \bibinfo
  {author} {\bibfnamefont {L.}~\bibnamefont {Li}}, \bibinfo {author}
  {\bibfnamefont {Y.-A.}\ \bibnamefont {Chen}}, \bibinfo {author}
  {\bibfnamefont {N.-L.}\ \bibnamefont {Liu}}, \bibinfo {author} {\bibfnamefont
  {C.-Y.}\ \bibnamefont {Lu}}, \ and\ \bibinfo {author} {\bibfnamefont {J.-W.}\
  \bibnamefont {Pan}},\ }\href {\doibase 10.1103/PhysRevLett.121.250505}
  {\bibfield  {journal} {\bibinfo  {journal} {Phys. Rev. Lett.}\ }\textbf
  {\bibinfo {volume} {121}},\ \bibinfo {pages} {250505} (\bibinfo {year}
  {2018})}\BibitemShut {NoStop}%
\bibitem [{\citenamefont {Xia}\ \emph {et~al.}(2018)\citenamefont {Xia},
  \citenamefont {Nori},\ and\ \citenamefont {Xiao}}]{PhysRevLett.121.203602}%
  \BibitemOpen
  \bibfield  {author} {\bibinfo {author} {\bibfnamefont {K.}~\bibnamefont
  {Xia}}, \bibinfo {author} {\bibfnamefont {F.}~\bibnamefont {Nori}}, \ and\
  \bibinfo {author} {\bibfnamefont {M.}~\bibnamefont {Xiao}},\ }\href {\doibase
  10.1103/PhysRevLett.121.203602} {\bibfield  {journal} {\bibinfo  {journal}
  {Phys. Rev. Lett.}\ }\textbf {\bibinfo {volume} {121}},\ \bibinfo {pages}
  {203602} (\bibinfo {year} {2018})}\BibitemShut {NoStop}%
\bibitem [{\citenamefont {Li}\ \emph {et~al.}(2020{\natexlab{a}})\citenamefont
  {Li}, \citenamefont {Liu}, \citenamefont {Ren}, \citenamefont {Wang},
  \citenamefont {Su}, \citenamefont {Chen}, \citenamefont {Chu}, \citenamefont
  {Kuo}, \citenamefont {Liu}, \citenamefont {Zang}, \citenamefont {Guo},
  \citenamefont {Zhang}, \citenamefont {Wang}, \citenamefont {Zhu},\ and\
  \citenamefont {Tsai}}]{science.aba9779}%
  \BibitemOpen
  \bibfield  {author} {\bibinfo {author} {\bibfnamefont {L.}~\bibnamefont
  {Li}}, \bibinfo {author} {\bibfnamefont {Z.}~\bibnamefont {Liu}}, \bibinfo
  {author} {\bibfnamefont {X.}~\bibnamefont {Ren}}, \bibinfo {author}
  {\bibfnamefont {S.}~\bibnamefont {Wang}}, \bibinfo {author} {\bibfnamefont
  {V.-C.}\ \bibnamefont {Su}}, \bibinfo {author} {\bibfnamefont {M.-K.}\
  \bibnamefont {Chen}}, \bibinfo {author} {\bibfnamefont {C.~H.}\ \bibnamefont
  {Chu}}, \bibinfo {author} {\bibfnamefont {H.~Y.}\ \bibnamefont {Kuo}},
  \bibinfo {author} {\bibfnamefont {B.}~\bibnamefont {Liu}}, \bibinfo {author}
  {\bibfnamefont {W.}~\bibnamefont {Zang}}, \bibinfo {author} {\bibfnamefont
  {G.}~\bibnamefont {Guo}}, \bibinfo {author} {\bibfnamefont {L.}~\bibnamefont
  {Zhang}}, \bibinfo {author} {\bibfnamefont {Z.}~\bibnamefont {Wang}},
  \bibinfo {author} {\bibfnamefont {S.}~\bibnamefont {Zhu}}, \ and\ \bibinfo
  {author} {\bibfnamefont {D.~P.}\ \bibnamefont {Tsai}},\ }\href {\doibase
  10.1126/science.aba9779} {\bibfield  {journal} {\bibinfo  {journal}
  {Science}\ }\textbf {\bibinfo {volume} {368}},\ \bibinfo {pages} {1487}
  (\bibinfo {year} {2020}{\natexlab{a}})}\BibitemShut {NoStop}%
\bibitem [{\citenamefont {Imamo\={g}lu}\ \emph {et~al.}(1997)\citenamefont
  {Imamo\={g}lu}, \citenamefont {Schmidt}, \citenamefont {Woods},\ and\
  \citenamefont {Deutsch}}]{PhysRevLett.79.1467}%
  \BibitemOpen
  \bibfield  {author} {\bibinfo {author} {\bibfnamefont {A.}~\bibnamefont
  {Imamo\={g}lu}}, \bibinfo {author} {\bibfnamefont {H.}~\bibnamefont
  {Schmidt}}, \bibinfo {author} {\bibfnamefont {G.}~\bibnamefont {Woods}}, \
  and\ \bibinfo {author} {\bibfnamefont {M.}~\bibnamefont {Deutsch}},\ }\href
  {\doibase 10.1103/PhysRevLett.79.1467} {\bibfield  {journal} {\bibinfo
  {journal} {Phys. Rev. Lett.}\ }\textbf {\bibinfo {volume} {79}},\ \bibinfo
  {pages} {1467} (\bibinfo {year} {1997})}\BibitemShut {NoStop}%
\bibitem [{\citenamefont {Sharping}\ \emph {et~al.}(2006)\citenamefont
  {Sharping}, \citenamefont {Lee}, \citenamefont {Foster}, \citenamefont
  {Turner}, \citenamefont {Schmidt}, \citenamefont {Lipson}, \citenamefont
  {Gaeta},\ and\ \citenamefont {Kumar}}]{OE.14.012388}%
  \BibitemOpen
  \bibfield  {author} {\bibinfo {author} {\bibfnamefont {J.~E.}\ \bibnamefont
  {Sharping}}, \bibinfo {author} {\bibfnamefont {K.~F.}\ \bibnamefont {Lee}},
  \bibinfo {author} {\bibfnamefont {M.~A.}\ \bibnamefont {Foster}}, \bibinfo
  {author} {\bibfnamefont {A.~C.}\ \bibnamefont {Turner}}, \bibinfo {author}
  {\bibfnamefont {B.~S.}\ \bibnamefont {Schmidt}}, \bibinfo {author}
  {\bibfnamefont {M.}~\bibnamefont {Lipson}}, \bibinfo {author} {\bibfnamefont
  {A.~L.}\ \bibnamefont {Gaeta}}, \ and\ \bibinfo {author} {\bibfnamefont
  {P.}~\bibnamefont {Kumar}},\ }\href {\doibase 10.1364/OE.14.012388}
  {\bibfield  {journal} {\bibinfo  {journal} {Opt. Express}\ }\textbf {\bibinfo
  {volume} {14}},\ \bibinfo {pages} {12388} (\bibinfo {year}
  {2006})}\BibitemShut {NoStop}%
\bibitem [{\citenamefont {Yuan}\ \emph {et~al.}(2011)\citenamefont {Yuan},
  \citenamefont {Chen}, \citenamefont {Ou},\ and\ \citenamefont
  {Zhang}}]{PhysRevA.83.054302}%
  \BibitemOpen
  \bibfield  {author} {\bibinfo {author} {\bibfnamefont {C.-H.}\ \bibnamefont
  {Yuan}}, \bibinfo {author} {\bibfnamefont {L.~Q.}\ \bibnamefont {Chen}},
  \bibinfo {author} {\bibfnamefont {Z.~Y.}\ \bibnamefont {Ou}}, \ and\ \bibinfo
  {author} {\bibfnamefont {W.}~\bibnamefont {Zhang}},\ }\href {\doibase
  10.1103/PhysRevA.83.054302} {\bibfield  {journal} {\bibinfo  {journal} {Phys.
  Rev. A}\ }\textbf {\bibinfo {volume} {83}},\ \bibinfo {pages} {054302}
  (\bibinfo {year} {2011})}\BibitemShut {NoStop}%
\bibitem [{\citenamefont {Silverstone}\ \emph {et~al.}(2014)\citenamefont
  {Silverstone}, \citenamefont {Bonneau}, \citenamefont {Ohira}, \citenamefont
  {Suzuki}, \citenamefont {Yoshida}, \citenamefont {Iizuka}, \citenamefont
  {Ezaki}, \citenamefont {Natarajan}, \citenamefont {Tanner}, \citenamefont
  {Hadfield}, \citenamefont {Zwiller}, \citenamefont {Marshall}, \citenamefont
  {Rarity}, \citenamefont {O'Brien},\ and\ \citenamefont
  {Thompson}}]{nphoton.2013.339}%
  \BibitemOpen
  \bibfield  {author} {\bibinfo {author} {\bibfnamefont {J.~W.}\ \bibnamefont
  {Silverstone}}, \bibinfo {author} {\bibfnamefont {D.}~\bibnamefont
  {Bonneau}}, \bibinfo {author} {\bibfnamefont {K.}~\bibnamefont {Ohira}},
  \bibinfo {author} {\bibfnamefont {N.}~\bibnamefont {Suzuki}}, \bibinfo
  {author} {\bibfnamefont {H.}~\bibnamefont {Yoshida}}, \bibinfo {author}
  {\bibfnamefont {N.}~\bibnamefont {Iizuka}}, \bibinfo {author} {\bibfnamefont
  {M.}~\bibnamefont {Ezaki}}, \bibinfo {author} {\bibfnamefont {C.~M.}\
  \bibnamefont {Natarajan}}, \bibinfo {author} {\bibfnamefont {M.~G.}\
  \bibnamefont {Tanner}}, \bibinfo {author} {\bibfnamefont {R.~H.}\
  \bibnamefont {Hadfield}}, \bibinfo {author} {\bibfnamefont {V.}~\bibnamefont
  {Zwiller}}, \bibinfo {author} {\bibfnamefont {G.~D.}\ \bibnamefont
  {Marshall}}, \bibinfo {author} {\bibfnamefont {J.~G.}\ \bibnamefont
  {Rarity}}, \bibinfo {author} {\bibfnamefont {J.~L.}\ \bibnamefont {O'Brien}},
  \ and\ \bibinfo {author} {\bibfnamefont {M.~G.}\ \bibnamefont {Thompson}},\
  }\href {\doibase 10.1038/nphoton.2013.339} {\bibfield  {journal} {\bibinfo
  {journal} {Nat. Photonics}\ }\textbf {\bibinfo {volume} {8}},\ \bibinfo
  {pages} {104} (\bibinfo {year} {2014})}\BibitemShut {NoStop}%
\bibitem [{\citenamefont {Caspani}\ \emph {et~al.}(2017)\citenamefont
  {Caspani}, \citenamefont {Xiong}, \citenamefont {Eggleton}, \citenamefont
  {Bajoni}, \citenamefont {Liscidini}, \citenamefont {Galli}, \citenamefont
  {Morandotti},\ and\ \citenamefont {Moss}}]{lsa.2017.100}%
  \BibitemOpen
  \bibfield  {author} {\bibinfo {author} {\bibfnamefont {L.}~\bibnamefont
  {Caspani}}, \bibinfo {author} {\bibfnamefont {C.}~\bibnamefont {Xiong}},
  \bibinfo {author} {\bibfnamefont {B.~J.}\ \bibnamefont {Eggleton}}, \bibinfo
  {author} {\bibfnamefont {D.}~\bibnamefont {Bajoni}}, \bibinfo {author}
  {\bibfnamefont {M.}~\bibnamefont {Liscidini}}, \bibinfo {author}
  {\bibfnamefont {M.}~\bibnamefont {Galli}}, \bibinfo {author} {\bibfnamefont
  {R.}~\bibnamefont {Morandotti}}, \ and\ \bibinfo {author} {\bibfnamefont
  {D.~J.}\ \bibnamefont {Moss}},\ }\href {\doibase 10.1038/lsa.2017.100}
  {\bibfield  {journal} {\bibinfo  {journal} {Light: Sci. Appl.}\ }\textbf
  {\bibinfo {volume} {6}},\ \bibinfo {pages} {e17100} (\bibinfo {year}
  {2017})}\BibitemShut {NoStop}%
\bibitem [{\citenamefont {Houck}\ \emph {et~al.}(2007)\citenamefont {Houck},
  \citenamefont {Schuster}, \citenamefont {Gambetta}, \citenamefont {Schreier},
  \citenamefont {Johnson}, \citenamefont {Chow}, \citenamefont {Frunzio},
  \citenamefont {Majer}, \citenamefont {Devoret}, \citenamefont {Girvin},\ and\
  \citenamefont {Schoelkopf}}]{nature06126}%
  \BibitemOpen
  \bibfield  {author} {\bibinfo {author} {\bibfnamefont {A.~A.}\ \bibnamefont
  {Houck}}, \bibinfo {author} {\bibfnamefont {D.~I.}\ \bibnamefont {Schuster}},
  \bibinfo {author} {\bibfnamefont {J.~M.}\ \bibnamefont {Gambetta}}, \bibinfo
  {author} {\bibfnamefont {J.~A.}\ \bibnamefont {Schreier}}, \bibinfo {author}
  {\bibfnamefont {B.~R.}\ \bibnamefont {Johnson}}, \bibinfo {author}
  {\bibfnamefont {J.~M.}\ \bibnamefont {Chow}}, \bibinfo {author}
  {\bibfnamefont {L.}~\bibnamefont {Frunzio}}, \bibinfo {author} {\bibfnamefont
  {J.}~\bibnamefont {Majer}}, \bibinfo {author} {\bibfnamefont {M.~H.}\
  \bibnamefont {Devoret}}, \bibinfo {author} {\bibfnamefont {S.~M.}\
  \bibnamefont {Girvin}}, \ and\ \bibinfo {author} {\bibfnamefont {R.~J.}\
  \bibnamefont {Schoelkopf}},\ }\href {\doibase 10.1038/nature06126} {\bibfield
   {journal} {\bibinfo  {journal} {Nature (London)}\ }\textbf {\bibinfo
  {volume} {449}},\ \bibinfo {pages} {328} (\bibinfo {year}
  {2007})}\BibitemShut {NoStop}%
\bibitem [{\citenamefont {Lang}\ \emph {et~al.}(2013)\citenamefont {Lang},
  \citenamefont {Eichler}, \citenamefont {Steffen}, \citenamefont {Fink},
  \citenamefont {Woolley}, \citenamefont {Blais},\ and\ \citenamefont
  {Wallraff}}]{nphys2612}%
  \BibitemOpen
  \bibfield  {author} {\bibinfo {author} {\bibfnamefont {C.}~\bibnamefont
  {Lang}}, \bibinfo {author} {\bibfnamefont {C.}~\bibnamefont {Eichler}},
  \bibinfo {author} {\bibfnamefont {L.}~\bibnamefont {Steffen}}, \bibinfo
  {author} {\bibfnamefont {J.~M.}\ \bibnamefont {Fink}}, \bibinfo {author}
  {\bibfnamefont {M.~J.}\ \bibnamefont {Woolley}}, \bibinfo {author}
  {\bibfnamefont {A.}~\bibnamefont {Blais}}, \ and\ \bibinfo {author}
  {\bibfnamefont {A.}~\bibnamefont {Wallraff}},\ }\href {\doibase
  10.1038/nphys2612} {\bibfield  {journal} {\bibinfo  {journal} {Nat. Phys.}\
  }\textbf {\bibinfo {volume} {9}},\ \bibinfo {pages} {345} (\bibinfo {year}
  {2013})}\BibitemShut {NoStop}%
\bibitem [{\citenamefont {Ding}\ \emph {et~al.}(2016)\citenamefont {Ding},
  \citenamefont {He}, \citenamefont {Duan}, \citenamefont {Gregersen},
  \citenamefont {Chen}, \citenamefont {Unsleber}, \citenamefont {Maier},
  \citenamefont {Schneider}, \citenamefont {Kamp}, \citenamefont {H\"ofling},
  \citenamefont {Lu},\ and\ \citenamefont {Pan}}]{PhysRevLett.116.020401}%
  \BibitemOpen
  \bibfield  {author} {\bibinfo {author} {\bibfnamefont {X.}~\bibnamefont
  {Ding}}, \bibinfo {author} {\bibfnamefont {Y.}~\bibnamefont {He}}, \bibinfo
  {author} {\bibfnamefont {Z.-C.}\ \bibnamefont {Duan}}, \bibinfo {author}
  {\bibfnamefont {N.}~\bibnamefont {Gregersen}}, \bibinfo {author}
  {\bibfnamefont {M.-C.}\ \bibnamefont {Chen}}, \bibinfo {author}
  {\bibfnamefont {S.}~\bibnamefont {Unsleber}}, \bibinfo {author}
  {\bibfnamefont {S.}~\bibnamefont {Maier}}, \bibinfo {author} {\bibfnamefont
  {C.}~\bibnamefont {Schneider}}, \bibinfo {author} {\bibfnamefont
  {M.}~\bibnamefont {Kamp}}, \bibinfo {author} {\bibfnamefont {S.}~\bibnamefont
  {H\"ofling}}, \bibinfo {author} {\bibfnamefont {C.-Y.}\ \bibnamefont {Lu}}, \
  and\ \bibinfo {author} {\bibfnamefont {J.-W.}\ \bibnamefont {Pan}},\ }\href
  {\doibase 10.1103/PhysRevLett.116.020401} {\bibfield  {journal} {\bibinfo
  {journal} {Phys. Rev. Lett.}\ }\textbf {\bibinfo {volume} {116}},\ \bibinfo
  {pages} {020401} (\bibinfo {year} {2016})}\BibitemShut {NoStop}%
\bibitem [{\citenamefont {Somaschi}\ \emph {et~al.}(2016)\citenamefont
  {Somaschi}, \citenamefont {Giesz}, \citenamefont {De~Santis}, \citenamefont
  {Loredo}, \citenamefont {Almeida}, \citenamefont {Hornecker}, \citenamefont
  {Portalupi}, \citenamefont {Grange}, \citenamefont {Ant\'{o}n}, \citenamefont
  {Demory}, \citenamefont {G\'{o}mez}, \citenamefont {Sagnes}, \citenamefont
  {Lanzillotti-Kimura}, \citenamefont {Lema\'{\i}tre}, \citenamefont
  {Auffeves}, \citenamefont {White}, \citenamefont {Lanco},\ and\ \citenamefont
  {Senellart}}]{nphoton.2016.23}%
  \BibitemOpen
  \bibfield  {author} {\bibinfo {author} {\bibfnamefont {N.}~\bibnamefont
  {Somaschi}}, \bibinfo {author} {\bibfnamefont {V.}~\bibnamefont {Giesz}},
  \bibinfo {author} {\bibfnamefont {L.}~\bibnamefont {De~Santis}}, \bibinfo
  {author} {\bibfnamefont {J.~C.}\ \bibnamefont {Loredo}}, \bibinfo {author}
  {\bibfnamefont {M.~P.}\ \bibnamefont {Almeida}}, \bibinfo {author}
  {\bibfnamefont {G.}~\bibnamefont {Hornecker}}, \bibinfo {author}
  {\bibfnamefont {S.~L.}\ \bibnamefont {Portalupi}}, \bibinfo {author}
  {\bibfnamefont {T.}~\bibnamefont {Grange}}, \bibinfo {author} {\bibfnamefont
  {C.}~\bibnamefont {Ant\'{o}n}}, \bibinfo {author} {\bibfnamefont
  {J.}~\bibnamefont {Demory}}, \bibinfo {author} {\bibfnamefont
  {C.}~\bibnamefont {G\'{o}mez}}, \bibinfo {author} {\bibfnamefont
  {I.}~\bibnamefont {Sagnes}}, \bibinfo {author} {\bibfnamefont {N.~D.}\
  \bibnamefont {Lanzillotti-Kimura}}, \bibinfo {author} {\bibfnamefont
  {A.}~\bibnamefont {Lema\'{\i}tre}}, \bibinfo {author} {\bibfnamefont
  {A.}~\bibnamefont {Auffeves}}, \bibinfo {author} {\bibfnamefont {A.~G.}\
  \bibnamefont {White}}, \bibinfo {author} {\bibfnamefont {L.}~\bibnamefont
  {Lanco}}, \ and\ \bibinfo {author} {\bibfnamefont {P.}~\bibnamefont
  {Senellart}},\ }\href {\doibase 10.1038/nphoton.2016.23} {\bibfield
  {journal} {\bibinfo  {journal} {Nat. Photonics}\ }\textbf {\bibinfo {volume}
  {10}},\ \bibinfo {pages} {340} (\bibinfo {year} {2016})}\BibitemShut
  {NoStop}%
\bibitem [{\citenamefont {Wang}\ \emph {et~al.}(2019)\citenamefont {Wang},
  \citenamefont {He}, \citenamefont {Chung}, \citenamefont {Hu}, \citenamefont
  {Yu}, \citenamefont {Chen}, \citenamefont {Ding}, \citenamefont {Chen},
  \citenamefont {Qin}, \citenamefont {Yang}, \citenamefont {Liu}, \citenamefont
  {Duan}, \citenamefont {Li}, \citenamefont {Gerhardt}, \citenamefont
  {Winkler}, \citenamefont {Jurkat}, \citenamefont {Wang}, \citenamefont
  {Gregersen}, \citenamefont {Huo}, \citenamefont {Dai}, \citenamefont {Yu},
  \citenamefont {H\"ofling}, \citenamefont {Lu},\ and\ \citenamefont
  {Pan}}]{nphoton.2019.10.1038}%
  \BibitemOpen
  \bibfield  {author} {\bibinfo {author} {\bibfnamefont {H.}~\bibnamefont
  {Wang}}, \bibinfo {author} {\bibfnamefont {Y.-M.}\ \bibnamefont {He}},
  \bibinfo {author} {\bibfnamefont {T.~H.}\ \bibnamefont {Chung}}, \bibinfo
  {author} {\bibfnamefont {H.}~\bibnamefont {Hu}}, \bibinfo {author}
  {\bibfnamefont {Y.}~\bibnamefont {Yu}}, \bibinfo {author} {\bibfnamefont
  {S.}~\bibnamefont {Chen}}, \bibinfo {author} {\bibfnamefont {X.}~\bibnamefont
  {Ding}}, \bibinfo {author} {\bibfnamefont {M.~C.}\ \bibnamefont {Chen}},
  \bibinfo {author} {\bibfnamefont {J.}~\bibnamefont {Qin}}, \bibinfo {author}
  {\bibfnamefont {X.}~\bibnamefont {Yang}}, \bibinfo {author} {\bibfnamefont
  {R.-Z.}\ \bibnamefont {Liu}}, \bibinfo {author} {\bibfnamefont {Z.~C.}\
  \bibnamefont {Duan}}, \bibinfo {author} {\bibfnamefont {J.~P.}\ \bibnamefont
  {Li}}, \bibinfo {author} {\bibfnamefont {S.}~\bibnamefont {Gerhardt}},
  \bibinfo {author} {\bibfnamefont {K.}~\bibnamefont {Winkler}}, \bibinfo
  {author} {\bibfnamefont {J.}~\bibnamefont {Jurkat}}, \bibinfo {author}
  {\bibfnamefont {L.-J.}\ \bibnamefont {Wang}}, \bibinfo {author}
  {\bibfnamefont {N.}~\bibnamefont {Gregersen}}, \bibinfo {author}
  {\bibfnamefont {Y.-H.}\ \bibnamefont {Huo}}, \bibinfo {author} {\bibfnamefont
  {Q.}~\bibnamefont {Dai}}, \bibinfo {author} {\bibfnamefont {S.}~\bibnamefont
  {Yu}}, \bibinfo {author} {\bibfnamefont {S.}~\bibnamefont {H\"ofling}},
  \bibinfo {author} {\bibfnamefont {C.-Y.}\ \bibnamefont {Lu}}, \ and\ \bibinfo
  {author} {\bibfnamefont {J.-W.}\ \bibnamefont {Pan}},\ }\href {\doibase
  10.1038/s41566-019-0494-3} {\bibfield  {journal} {\bibinfo  {journal} {Nat.
  Photonics}\ }\textbf {\bibinfo {volume} {13}},\ \bibinfo {pages} {770}
  (\bibinfo {year} {2019})}\BibitemShut {NoStop}%
\bibitem [{\citenamefont {He}\ \emph {et~al.}(2019)\citenamefont {He},
  \citenamefont {Wang}, \citenamefont {Wang}, \citenamefont {Chen},
  \citenamefont {Ding}, \citenamefont {Qin}, \citenamefont {Duan},
  \citenamefont {Chen}, \citenamefont {Li}, \citenamefont {Liu}, \citenamefont
  {Schneider}, \citenamefont {Atat\"{u}re}, \citenamefont {H\"{o}fling},
  \citenamefont {Lu},\ and\ \citenamefont {Pan}}]{nphysics.10.1038.941}%
  \BibitemOpen
  \bibfield  {author} {\bibinfo {author} {\bibfnamefont {Y.-M.}\ \bibnamefont
  {He}}, \bibinfo {author} {\bibfnamefont {H.}~\bibnamefont {Wang}}, \bibinfo
  {author} {\bibfnamefont {C.}~\bibnamefont {Wang}}, \bibinfo {author}
  {\bibfnamefont {M.~C.}\ \bibnamefont {Chen}}, \bibinfo {author}
  {\bibfnamefont {X.}~\bibnamefont {Ding}}, \bibinfo {author} {\bibfnamefont
  {J.}~\bibnamefont {Qin}}, \bibinfo {author} {\bibfnamefont {Z.~C.}\
  \bibnamefont {Duan}}, \bibinfo {author} {\bibfnamefont {S.}~\bibnamefont
  {Chen}}, \bibinfo {author} {\bibfnamefont {J.~P.}\ \bibnamefont {Li}},
  \bibinfo {author} {\bibfnamefont {R.-Z.}\ \bibnamefont {Liu}}, \bibinfo
  {author} {\bibfnamefont {C.}~\bibnamefont {Schneider}}, \bibinfo {author}
  {\bibfnamefont {M.}~\bibnamefont {Atat\"{u}re}}, \bibinfo {author}
  {\bibfnamefont {S.}~\bibnamefont {H\"{o}fling}}, \bibinfo {author}
  {\bibfnamefont {C.-Y.}\ \bibnamefont {Lu}}, \ and\ \bibinfo {author}
  {\bibfnamefont {J.-W.}\ \bibnamefont {Pan}},\ }\href {\doibase
  10.1038/s41567-019-0585-6} {\bibfield  {journal} {\bibinfo  {journal} {Nat.
  Phys.}\ }\textbf {\bibinfo {volume} {15}},\ \bibinfo {pages} {941} (\bibinfo
  {year} {2019})}\BibitemShut {NoStop}%
\bibitem [{\citenamefont {Koong}\ \emph {et~al.}(2021)\citenamefont {Koong},
  \citenamefont {Scerri}, \citenamefont {Rambach}, \citenamefont {Cygorek},
  \citenamefont {Brotons-Gisbert}, \citenamefont {Picard}, \citenamefont {Ma},
  \citenamefont {Park}, \citenamefont {Song}, \citenamefont {Gauger},\ and\
  \citenamefont {Gerardot}}]{PhysRevLett.126.047403}%
  \BibitemOpen
  \bibfield  {author} {\bibinfo {author} {\bibfnamefont {Z.~X.}\ \bibnamefont
  {Koong}}, \bibinfo {author} {\bibfnamefont {E.}~\bibnamefont {Scerri}},
  \bibinfo {author} {\bibfnamefont {M.}~\bibnamefont {Rambach}}, \bibinfo
  {author} {\bibfnamefont {M.}~\bibnamefont {Cygorek}}, \bibinfo {author}
  {\bibfnamefont {M.}~\bibnamefont {Brotons-Gisbert}}, \bibinfo {author}
  {\bibfnamefont {R.}~\bibnamefont {Picard}}, \bibinfo {author} {\bibfnamefont
  {Y.}~\bibnamefont {Ma}}, \bibinfo {author} {\bibfnamefont {S.~I.}\
  \bibnamefont {Park}}, \bibinfo {author} {\bibfnamefont {J.~D.}\ \bibnamefont
  {Song}}, \bibinfo {author} {\bibfnamefont {E.~M.}\ \bibnamefont {Gauger}}, \
  and\ \bibinfo {author} {\bibfnamefont {B.~D.}\ \bibnamefont {Gerardot}},\
  }\href {\doibase 10.1103/PhysRevLett.126.047403} {\bibfield  {journal}
  {\bibinfo  {journal} {Phys. Rev. Lett.}\ }\textbf {\bibinfo {volume} {126}},\
  \bibinfo {pages} {047403} (\bibinfo {year} {2021})}\BibitemShut {NoStop}%
\bibitem [{\citenamefont {Ou}\ and\ \citenamefont
  {Lu}(1999)}]{PhysRevLett.83.2556}%
  \BibitemOpen
  \bibfield  {author} {\bibinfo {author} {\bibfnamefont {Z.~Y.}\ \bibnamefont
  {Ou}}\ and\ \bibinfo {author} {\bibfnamefont {Y.~J.}\ \bibnamefont {Lu}},\
  }\href {\doibase 10.1103/PhysRevLett.83.2556} {\bibfield  {journal} {\bibinfo
   {journal} {Phys. Rev. Lett.}\ }\textbf {\bibinfo {volume} {83}},\ \bibinfo
  {pages} {2556} (\bibinfo {year} {1999})}\BibitemShut {NoStop}%
\bibitem [{\citenamefont {Zhang}\ \emph {et~al.}(2011)\citenamefont {Zhang},
  \citenamefont {Jin}, \citenamefont {Yang}, \citenamefont {Dai}, \citenamefont
  {Yang}, \citenamefont {Zhao}, \citenamefont {Rui}, \citenamefont {He},
  \citenamefont {Jiang}, \citenamefont {Yang}, \citenamefont {Pan},
  \citenamefont {Yuan}, \citenamefont {Deng}, \citenamefont {Chen},
  \citenamefont {Bao}, \citenamefont {Chen}, \citenamefont {Zhao},\ and\
  \citenamefont {Pan}}]{nphoton.2011.213}%
  \BibitemOpen
  \bibfield  {author} {\bibinfo {author} {\bibfnamefont {H.}~\bibnamefont
  {Zhang}}, \bibinfo {author} {\bibfnamefont {X.-M.}\ \bibnamefont {Jin}},
  \bibinfo {author} {\bibfnamefont {J.}~\bibnamefont {Yang}}, \bibinfo {author}
  {\bibfnamefont {H.-N.}\ \bibnamefont {Dai}}, \bibinfo {author} {\bibfnamefont
  {S.-J.}\ \bibnamefont {Yang}}, \bibinfo {author} {\bibfnamefont {T.-M.}\
  \bibnamefont {Zhao}}, \bibinfo {author} {\bibfnamefont {J.}~\bibnamefont
  {Rui}}, \bibinfo {author} {\bibfnamefont {Y.}~\bibnamefont {He}}, \bibinfo
  {author} {\bibfnamefont {X.}~\bibnamefont {Jiang}}, \bibinfo {author}
  {\bibfnamefont {F.}~\bibnamefont {Yang}}, \bibinfo {author} {\bibfnamefont
  {G.-S.}\ \bibnamefont {Pan}}, \bibinfo {author} {\bibfnamefont {Z.-S.}\
  \bibnamefont {Yuan}}, \bibinfo {author} {\bibfnamefont {Y.}~\bibnamefont
  {Deng}}, \bibinfo {author} {\bibfnamefont {Z.-B.}\ \bibnamefont {Chen}},
  \bibinfo {author} {\bibfnamefont {X.-H.}\ \bibnamefont {Bao}}, \bibinfo
  {author} {\bibfnamefont {S.}~\bibnamefont {Chen}}, \bibinfo {author}
  {\bibfnamefont {B.}~\bibnamefont {Zhao}}, \ and\ \bibinfo {author}
  {\bibfnamefont {J.-W.}\ \bibnamefont {Pan}},\ }\href {\doibase
  10.1038/nphoton.2011.213} {\bibfield  {journal} {\bibinfo  {journal} {Nat.
  Photonics}\ }\textbf {\bibinfo {volume} {5}},\ \bibinfo {pages} {628}
  (\bibinfo {year} {2011})}\BibitemShut {NoStop}%
\bibitem [{\citenamefont {Guo}\ \emph {et~al.}(2017)\citenamefont {Guo},
  \citenamefont {Zou}, \citenamefont {Schuck}, \citenamefont {Jung},
  \citenamefont {Cheng},\ and\ \citenamefont {Tang}}]{lsa.2016.249}%
  \BibitemOpen
  \bibfield  {author} {\bibinfo {author} {\bibfnamefont {X.}~\bibnamefont
  {Guo}}, \bibinfo {author} {\bibfnamefont {C.-l.}\ \bibnamefont {Zou}},
  \bibinfo {author} {\bibfnamefont {C.}~\bibnamefont {Schuck}}, \bibinfo
  {author} {\bibfnamefont {H.}~\bibnamefont {Jung}}, \bibinfo {author}
  {\bibfnamefont {R.}~\bibnamefont {Cheng}}, \ and\ \bibinfo {author}
  {\bibfnamefont {H.~X.}\ \bibnamefont {Tang}},\ }\href {\doibase
  10.1038/lsa.2016.249} {\bibfield  {journal} {\bibinfo  {journal} {Light: Sci.
  Appl.}\ }\textbf {\bibinfo {volume} {6}},\ \bibinfo {pages} {e16249}
  (\bibinfo {year} {2017})}\BibitemShut {NoStop}%
\bibitem [{\citenamefont {Ma}\ \emph {et~al.}(2020)\citenamefont {Ma},
  \citenamefont {Chen}, \citenamefont {Li}, \citenamefont {Tang}, \citenamefont
  {Sua}, \citenamefont {Fan},\ and\ \citenamefont
  {Huang}}]{PhysRevLett.125.263602}%
  \BibitemOpen
  \bibfield  {author} {\bibinfo {author} {\bibfnamefont {Z.}~\bibnamefont
  {Ma}}, \bibinfo {author} {\bibfnamefont {J.-Y.}\ \bibnamefont {Chen}},
  \bibinfo {author} {\bibfnamefont {Z.}~\bibnamefont {Li}}, \bibinfo {author}
  {\bibfnamefont {C.}~\bibnamefont {Tang}}, \bibinfo {author} {\bibfnamefont
  {Y.~M.}\ \bibnamefont {Sua}}, \bibinfo {author} {\bibfnamefont
  {H.}~\bibnamefont {Fan}}, \ and\ \bibinfo {author} {\bibfnamefont {Y.-P.}\
  \bibnamefont {Huang}},\ }\href {\doibase 10.1103/PhysRevLett.125.263602}
  {\bibfield  {journal} {\bibinfo  {journal} {Phys. Rev. Lett.}\ }\textbf
  {\bibinfo {volume} {125}},\ \bibinfo {pages} {263602} (\bibinfo {year}
  {2020})}\BibitemShut {NoStop}%
\bibitem [{\citenamefont {Collins}\ \emph {et~al.}(2013)\citenamefont
  {Collins}, \citenamefont {Xiong}, \citenamefont {Rey}, \citenamefont {Vo},
  \citenamefont {He}, \citenamefont {Shahnia}, \citenamefont {Reardon},
  \citenamefont {Krauss}, \citenamefont {Steel}, \citenamefont {Clark},\ and\
  \citenamefont {Eggleton}}]{ncomms3582}%
  \BibitemOpen
  \bibfield  {author} {\bibinfo {author} {\bibfnamefont {M.~J.}\ \bibnamefont
  {Collins}}, \bibinfo {author} {\bibfnamefont {C.}~\bibnamefont {Xiong}},
  \bibinfo {author} {\bibfnamefont {I.~H.}\ \bibnamefont {Rey}}, \bibinfo
  {author} {\bibfnamefont {T.~D.}\ \bibnamefont {Vo}}, \bibinfo {author}
  {\bibfnamefont {J.}~\bibnamefont {He}}, \bibinfo {author} {\bibfnamefont
  {S.}~\bibnamefont {Shahnia}}, \bibinfo {author} {\bibfnamefont
  {C.}~\bibnamefont {Reardon}}, \bibinfo {author} {\bibfnamefont {T.~F.}\
  \bibnamefont {Krauss}}, \bibinfo {author} {\bibfnamefont {M.~J.}\
  \bibnamefont {Steel}}, \bibinfo {author} {\bibfnamefont {A.~S.}\ \bibnamefont
  {Clark}}, \ and\ \bibinfo {author} {\bibfnamefont {B.~J.}\ \bibnamefont
  {Eggleton}},\ }\href {\doibase 10.1038/ncomms3582} {\bibfield  {journal}
  {\bibinfo  {journal} {Nat. Commun.}\ }\textbf {\bibinfo {volume} {4}},\
  \bibinfo {pages} {2582} (\bibinfo {year} {2013})}\BibitemShut {NoStop}%
\bibitem [{\citenamefont {Reim}\ \emph {et~al.}(2010)\citenamefont {Reim},
  \citenamefont {Nunn}, \citenamefont {Lorenz}, \citenamefont {Sussman},
  \citenamefont {Lee}, \citenamefont {Langford}, \citenamefont {Jaksch},\ and\
  \citenamefont {Walmsley}}]{nphoton.2010.30}%
  \BibitemOpen
  \bibfield  {author} {\bibinfo {author} {\bibfnamefont {K.~F.}\ \bibnamefont
  {Reim}}, \bibinfo {author} {\bibfnamefont {J.}~\bibnamefont {Nunn}}, \bibinfo
  {author} {\bibfnamefont {V.~O.}\ \bibnamefont {Lorenz}}, \bibinfo {author}
  {\bibfnamefont {B.~J.}\ \bibnamefont {Sussman}}, \bibinfo {author}
  {\bibfnamefont {K.~C.}\ \bibnamefont {Lee}}, \bibinfo {author} {\bibfnamefont
  {N.~K.}\ \bibnamefont {Langford}}, \bibinfo {author} {\bibfnamefont
  {D.}~\bibnamefont {Jaksch}}, \ and\ \bibinfo {author} {\bibfnamefont {I.~A.}\
  \bibnamefont {Walmsley}},\ }\href {\doibase 10.1038/nphoton.2010.30}
  {\bibfield  {journal} {\bibinfo  {journal} {Nat. Photonics}\ }\textbf
  {\bibinfo {volume} {4}},\ \bibinfo {pages} {218} (\bibinfo {year}
  {2010})}\BibitemShut {NoStop}%
\bibitem [{\citenamefont {Christ}\ and\ \citenamefont
  {Silberhorn}(2012)}]{PhysRevA.85.023829}%
  \BibitemOpen
  \bibfield  {author} {\bibinfo {author} {\bibfnamefont {A.}~\bibnamefont
  {Christ}}\ and\ \bibinfo {author} {\bibfnamefont {C.}~\bibnamefont
  {Silberhorn}},\ }\href {\doibase 10.1103/PhysRevA.85.023829} {\bibfield
  {journal} {\bibinfo  {journal} {Phys. Rev. A}\ }\textbf {\bibinfo {volume}
  {85}},\ \bibinfo {pages} {023829} (\bibinfo {year} {2012})}\BibitemShut
  {NoStop}%
\bibitem [{\citenamefont {Takeoka}\ \emph {et~al.}(2015)\citenamefont
  {Takeoka}, \citenamefont {Jin},\ and\ \citenamefont {Sasaki}}]{njp.043030}%
  \BibitemOpen
  \bibfield  {author} {\bibinfo {author} {\bibfnamefont {M.}~\bibnamefont
  {Takeoka}}, \bibinfo {author} {\bibfnamefont {R.-B.}\ \bibnamefont {Jin}}, \
  and\ \bibinfo {author} {\bibfnamefont {M.}~\bibnamefont {Sasaki}},\ }\href
  {\doibase 10.1088/1367-2630/17/4/043030} {\bibfield  {journal} {\bibinfo
  {journal} {New J. Phys.}\ }\textbf {\bibinfo {volume} {17}},\ \bibinfo
  {pages} {043030} (\bibinfo {year} {2015})}\BibitemShut {NoStop}%
\bibitem [{\citenamefont {Jennewein}\ \emph {et~al.}(2011)\citenamefont
  {Jennewein}, \citenamefont {Barbieri},\ and\ \citenamefont
  {White}}]{JMO.2010.546894}%
  \BibitemOpen
  \bibfield  {author} {\bibinfo {author} {\bibfnamefont {T.}~\bibnamefont
  {Jennewein}}, \bibinfo {author} {\bibfnamefont {M.}~\bibnamefont {Barbieri}},
  \ and\ \bibinfo {author} {\bibfnamefont {A.~G.}\ \bibnamefont {White}},\
  }\href {\doibase 10.1080/09500340.2010.546894} {\bibfield  {journal}
  {\bibinfo  {journal} {J. Mod. Opt.}\ }\textbf {\bibinfo {volume} {58}},\
  \bibinfo {pages} {276} (\bibinfo {year} {2011})}\BibitemShut {NoStop}%
\bibitem [{\citenamefont {Migdall}\ \emph {et~al.}(2002)\citenamefont
  {Migdall}, \citenamefont {Branning},\ and\ \citenamefont
  {Castelletto}}]{PhysRevA.66.053805}%
  \BibitemOpen
  \bibfield  {author} {\bibinfo {author} {\bibfnamefont {A.~L.}\ \bibnamefont
  {Migdall}}, \bibinfo {author} {\bibfnamefont {D.}~\bibnamefont {Branning}}, \
  and\ \bibinfo {author} {\bibfnamefont {S.}~\bibnamefont {Castelletto}},\
  }\href {\doibase 10.1103/PhysRevA.66.053805} {\bibfield  {journal} {\bibinfo
  {journal} {Phys. Rev. A}\ }\textbf {\bibinfo {volume} {66}},\ \bibinfo
  {pages} {053805} (\bibinfo {year} {2002})}\BibitemShut {NoStop}%
\bibitem [{\citenamefont {Pittman}\ \emph {et~al.}(2002)\citenamefont
  {Pittman}, \citenamefont {Jacobs},\ and\ \citenamefont
  {Franson}}]{PhysRevA.66.042303}%
  \BibitemOpen
  \bibfield  {author} {\bibinfo {author} {\bibfnamefont {T.~B.}\ \bibnamefont
  {Pittman}}, \bibinfo {author} {\bibfnamefont {B.~C.}\ \bibnamefont {Jacobs}},
  \ and\ \bibinfo {author} {\bibfnamefont {J.~D.}\ \bibnamefont {Franson}},\
  }\href {\doibase 10.1103/PhysRevA.66.042303} {\bibfield  {journal} {\bibinfo
  {journal} {Phys. Rev. A}\ }\textbf {\bibinfo {volume} {66}},\ \bibinfo
  {pages} {042303} (\bibinfo {year} {2002})}\BibitemShut {NoStop}%
\bibitem [{\citenamefont {Grimau~Puigibert}\ \emph {et~al.}(2017)\citenamefont
  {Grimau~Puigibert}, \citenamefont {Aguilar}, \citenamefont {Zhou},
  \citenamefont {Marsili}, \citenamefont {Shaw}, \citenamefont {Verma},
  \citenamefont {Nam}, \citenamefont {Oblak},\ and\ \citenamefont
  {Tittel}}]{PhysRevLett.119.083601}%
  \BibitemOpen
  \bibfield  {author} {\bibinfo {author} {\bibfnamefont {M.}~\bibnamefont
  {Grimau~Puigibert}}, \bibinfo {author} {\bibfnamefont {G.~H.}\ \bibnamefont
  {Aguilar}}, \bibinfo {author} {\bibfnamefont {Q.}~\bibnamefont {Zhou}},
  \bibinfo {author} {\bibfnamefont {F.}~\bibnamefont {Marsili}}, \bibinfo
  {author} {\bibfnamefont {M.~D.}\ \bibnamefont {Shaw}}, \bibinfo {author}
  {\bibfnamefont {V.~B.}\ \bibnamefont {Verma}}, \bibinfo {author}
  {\bibfnamefont {S.~W.}\ \bibnamefont {Nam}}, \bibinfo {author} {\bibfnamefont
  {D.}~\bibnamefont {Oblak}}, \ and\ \bibinfo {author} {\bibfnamefont
  {W.}~\bibnamefont {Tittel}},\ }\href {\doibase
  10.1103/PhysRevLett.119.083601} {\bibfield  {journal} {\bibinfo  {journal}
  {Phys. Rev. Lett.}\ }\textbf {\bibinfo {volume} {119}},\ \bibinfo {pages}
  {083601} (\bibinfo {year} {2017})}\BibitemShut {NoStop}%
\bibitem [{\citenamefont {Kaneda}\ and\ \citenamefont
  {Kwiat}(2019)}]{sciadv.aaw8586}%
  \BibitemOpen
  \bibfield  {author} {\bibinfo {author} {\bibfnamefont {F.}~\bibnamefont
  {Kaneda}}\ and\ \bibinfo {author} {\bibfnamefont {P.~G.}\ \bibnamefont
  {Kwiat}},\ }\href {\doibase 10.1126/sciadv.aaw8586} {\bibfield  {journal}
  {\bibinfo  {journal} {Sci. Adv.}\ }\textbf {\bibinfo {volume} {5}} (\bibinfo
  {year} {2019}),\ 10.1126/sciadv.aaw8586}\BibitemShut {NoStop}%
\bibitem [{\citenamefont {Neergaard-Nielsen}\ \emph {et~al.}(2007)\citenamefont
  {Neergaard-Nielsen}, \citenamefont {Nielsen}, \citenamefont {Takahashi},
  \citenamefont {Vistnes},\ and\ \citenamefont {Polzik}}]{OE.15.007940}%
  \BibitemOpen
  \bibfield  {author} {\bibinfo {author} {\bibfnamefont {J.~S.}\ \bibnamefont
  {Neergaard-Nielsen}}, \bibinfo {author} {\bibfnamefont {B.~M.}\ \bibnamefont
  {Nielsen}}, \bibinfo {author} {\bibfnamefont {H.}~\bibnamefont {Takahashi}},
  \bibinfo {author} {\bibfnamefont {A.~I.}\ \bibnamefont {Vistnes}}, \ and\
  \bibinfo {author} {\bibfnamefont {E.~S.}\ \bibnamefont {Polzik}},\ }\href
  {\doibase 10.1364/OE.15.007940} {\bibfield  {journal} {\bibinfo  {journal}
  {Opt. Express}\ }\textbf {\bibinfo {volume} {15}},\ \bibinfo {pages} {7940}
  (\bibinfo {year} {2007})}\BibitemShut {NoStop}%
\bibitem [{\citenamefont {Bao}\ \emph {et~al.}(2008)\citenamefont {Bao},
  \citenamefont {Qian}, \citenamefont {Yang}, \citenamefont {Zhang},
  \citenamefont {Chen}, \citenamefont {Yang},\ and\ \citenamefont
  {Pan}}]{PhysRevLett.101.190501}%
  \BibitemOpen
  \bibfield  {author} {\bibinfo {author} {\bibfnamefont {X.-H.}\ \bibnamefont
  {Bao}}, \bibinfo {author} {\bibfnamefont {Y.}~\bibnamefont {Qian}}, \bibinfo
  {author} {\bibfnamefont {J.}~\bibnamefont {Yang}}, \bibinfo {author}
  {\bibfnamefont {H.}~\bibnamefont {Zhang}}, \bibinfo {author} {\bibfnamefont
  {Z.-B.}\ \bibnamefont {Chen}}, \bibinfo {author} {\bibfnamefont
  {T.}~\bibnamefont {Yang}}, \ and\ \bibinfo {author} {\bibfnamefont {J.-W.}\
  \bibnamefont {Pan}},\ }\href {\doibase 10.1103/PhysRevLett.101.190501}
  {\bibfield  {journal} {\bibinfo  {journal} {Phys. Rev. Lett.}\ }\textbf
  {\bibinfo {volume} {101}},\ \bibinfo {pages} {190501} (\bibinfo {year}
  {2008})}\BibitemShut {NoStop}%
\bibitem [{\citenamefont {Dai}\ \emph {et~al.}(2012)\citenamefont {Dai},
  \citenamefont {Zhang}, \citenamefont {Yang}, \citenamefont {Zhao},
  \citenamefont {Rui}, \citenamefont {Deng}, \citenamefont {Li}, \citenamefont
  {Liu}, \citenamefont {Chen}, \citenamefont {Bao}, \citenamefont {Jin},
  \citenamefont {Zhao},\ and\ \citenamefont {Pan}}]{PhysRevLett.108.210501}%
  \BibitemOpen
  \bibfield  {author} {\bibinfo {author} {\bibfnamefont {H.-N.}\ \bibnamefont
  {Dai}}, \bibinfo {author} {\bibfnamefont {H.}~\bibnamefont {Zhang}}, \bibinfo
  {author} {\bibfnamefont {S.-J.}\ \bibnamefont {Yang}}, \bibinfo {author}
  {\bibfnamefont {T.-M.}\ \bibnamefont {Zhao}}, \bibinfo {author}
  {\bibfnamefont {J.}~\bibnamefont {Rui}}, \bibinfo {author} {\bibfnamefont
  {Y.-J.}\ \bibnamefont {Deng}}, \bibinfo {author} {\bibfnamefont
  {L.}~\bibnamefont {Li}}, \bibinfo {author} {\bibfnamefont {N.-L.}\
  \bibnamefont {Liu}}, \bibinfo {author} {\bibfnamefont {S.}~\bibnamefont
  {Chen}}, \bibinfo {author} {\bibfnamefont {X.-H.}\ \bibnamefont {Bao}},
  \bibinfo {author} {\bibfnamefont {X.-M.}\ \bibnamefont {Jin}}, \bibinfo
  {author} {\bibfnamefont {B.}~\bibnamefont {Zhao}}, \ and\ \bibinfo {author}
  {\bibfnamefont {J.-W.}\ \bibnamefont {Pan}},\ }\href {\doibase
  10.1103/PhysRevLett.108.210501} {\bibfield  {journal} {\bibinfo  {journal}
  {Phys. Rev. Lett.}\ }\textbf {\bibinfo {volume} {108}},\ \bibinfo {pages}
  {210501} (\bibinfo {year} {2012})}\BibitemShut {NoStop}%
\bibitem [{\citenamefont {Sheng}\ \emph {et~al.}(2011)\citenamefont {Sheng},
  \citenamefont {Yang}, \citenamefont {Wu},\ and\ \citenamefont
  {Xiao}}]{PhysRevA.84.053820}%
  \BibitemOpen
  \bibfield  {author} {\bibinfo {author} {\bibfnamefont {J.}~\bibnamefont
  {Sheng}}, \bibinfo {author} {\bibfnamefont {X.}~\bibnamefont {Yang}},
  \bibinfo {author} {\bibfnamefont {H.}~\bibnamefont {Wu}}, \ and\ \bibinfo
  {author} {\bibfnamefont {M.}~\bibnamefont {Xiao}},\ }\href {\doibase
  10.1103/PhysRevA.84.053820} {\bibfield  {journal} {\bibinfo  {journal} {Phys.
  Rev. A}\ }\textbf {\bibinfo {volume} {84}},\ \bibinfo {pages} {053820}
  (\bibinfo {year} {2011})}\BibitemShut {NoStop}%
\bibitem [{\citenamefont {Liu}\ \emph {et~al.}(2016)\citenamefont {Liu},
  \citenamefont {Chen}, \citenamefont {Chen}, \citenamefont {Lo}, \citenamefont
  {Tsai}, \citenamefont {Yu}, \citenamefont {Chen},\ and\ \citenamefont
  {Chen}}]{PhysRevLett.117.203601}%
  \BibitemOpen
  \bibfield  {author} {\bibinfo {author} {\bibfnamefont {Z.-Y.}\ \bibnamefont
  {Liu}}, \bibinfo {author} {\bibfnamefont {Y.-H.}\ \bibnamefont {Chen}},
  \bibinfo {author} {\bibfnamefont {Y.-C.}\ \bibnamefont {Chen}}, \bibinfo
  {author} {\bibfnamefont {H.-Y.}\ \bibnamefont {Lo}}, \bibinfo {author}
  {\bibfnamefont {P.-J.}\ \bibnamefont {Tsai}}, \bibinfo {author}
  {\bibfnamefont {I.~A.}\ \bibnamefont {Yu}}, \bibinfo {author} {\bibfnamefont
  {Y.-C.}\ \bibnamefont {Chen}}, \ and\ \bibinfo {author} {\bibfnamefont
  {Y.-F.}\ \bibnamefont {Chen}},\ }\href {\doibase
  10.1103/PhysRevLett.117.203601} {\bibfield  {journal} {\bibinfo  {journal}
  {Phys. Rev. Lett.}\ }\textbf {\bibinfo {volume} {117}},\ \bibinfo {pages}
  {203601} (\bibinfo {year} {2016})}\BibitemShut {NoStop}%
\bibitem [{\citenamefont {Sagona-Stophel}\ \emph {et~al.}(2020)\citenamefont
  {Sagona-Stophel}, \citenamefont {Shahrokhshahi}, \citenamefont {Jordaan},
  \citenamefont {Namazi},\ and\ \citenamefont
  {Figueroa}}]{PhysRevLett.125.243601}%
  \BibitemOpen
  \bibfield  {author} {\bibinfo {author} {\bibfnamefont {S.}~\bibnamefont
  {Sagona-Stophel}}, \bibinfo {author} {\bibfnamefont {R.}~\bibnamefont
  {Shahrokhshahi}}, \bibinfo {author} {\bibfnamefont {B.}~\bibnamefont
  {Jordaan}}, \bibinfo {author} {\bibfnamefont {M.}~\bibnamefont {Namazi}}, \
  and\ \bibinfo {author} {\bibfnamefont {E.}~\bibnamefont {Figueroa}},\ }\href
  {\doibase 10.1103/PhysRevLett.125.243601} {\bibfield  {journal} {\bibinfo
  {journal} {Phys. Rev. Lett.}\ }\textbf {\bibinfo {volume} {125}},\ \bibinfo
  {pages} {243601} (\bibinfo {year} {2020})}\BibitemShut {NoStop}%
\bibitem [{\citenamefont {Sangouard}\ \emph {et~al.}(2011)\citenamefont
  {Sangouard}, \citenamefont {Simon}, \citenamefont {de~Riedmatten},\ and\
  \citenamefont {Gisin}}]{RevModPhys.83.33}%
  \BibitemOpen
  \bibfield  {author} {\bibinfo {author} {\bibfnamefont {N.}~\bibnamefont
  {Sangouard}}, \bibinfo {author} {\bibfnamefont {C.}~\bibnamefont {Simon}},
  \bibinfo {author} {\bibfnamefont {H.}~\bibnamefont {de~Riedmatten}}, \ and\
  \bibinfo {author} {\bibfnamefont {N.}~\bibnamefont {Gisin}},\ }\href
  {\doibase 10.1103/RevModPhys.83.33} {\bibfield  {journal} {\bibinfo
  {journal} {Rev. Mod. Phys.}\ }\textbf {\bibinfo {volume} {83}},\ \bibinfo
  {pages} {33} (\bibinfo {year} {2011})}\BibitemShut {NoStop}%
\bibitem [{\citenamefont {Tang}\ \emph {et~al.}(2020)\citenamefont {Tang},
  \citenamefont {Wu}, \citenamefont {Wang}, \citenamefont {Sun}, \citenamefont
  {Tang}, \citenamefont {Zhang}, \citenamefont {Li}, \citenamefont {Lu},
  \citenamefont {Xiao},\ and\ \citenamefont {Xia}}]{PhysRevA.101.053802}%
  \BibitemOpen
  \bibfield  {author} {\bibinfo {author} {\bibfnamefont {J.}~\bibnamefont
  {Tang}}, \bibinfo {author} {\bibfnamefont {Y.}~\bibnamefont {Wu}}, \bibinfo
  {author} {\bibfnamefont {Z.}~\bibnamefont {Wang}}, \bibinfo {author}
  {\bibfnamefont {H.}~\bibnamefont {Sun}}, \bibinfo {author} {\bibfnamefont
  {L.}~\bibnamefont {Tang}}, \bibinfo {author} {\bibfnamefont {H.}~\bibnamefont
  {Zhang}}, \bibinfo {author} {\bibfnamefont {T.}~\bibnamefont {Li}}, \bibinfo
  {author} {\bibfnamefont {Y.}~\bibnamefont {Lu}}, \bibinfo {author}
  {\bibfnamefont {M.}~\bibnamefont {Xiao}}, \ and\ \bibinfo {author}
  {\bibfnamefont {K.}~\bibnamefont {Xia}},\ }\href {\doibase
  10.1103/PhysRevA.101.053802} {\bibfield  {journal} {\bibinfo  {journal}
  {Phys. Rev. A}\ }\textbf {\bibinfo {volume} {101}},\ \bibinfo {pages}
  {053802} (\bibinfo {year} {2020})}\BibitemShut {NoStop}%
\bibitem [{\citenamefont {Scully}\ \emph {et~al.}(1997)\citenamefont {Scully},
  \citenamefont {Zubairy} \emph {et~al.}}]{scully1997quantum}%
  \BibitemOpen
  \bibfield  {author} {\bibinfo {author} {\bibfnamefont {M.~O.}\ \bibnamefont
  {Scully}}, \bibinfo {author} {\bibfnamefont {M.~S.}\ \bibnamefont {Zubairy}},
   \emph {et~al.},\ }\href@noop {} {\emph {\bibinfo {title} {Quantum Optics}}}\
  (\bibinfo  {publisher} {Cambridge University Press},\ \bibinfo {year}
  {1997})\BibitemShut {NoStop}%
\bibitem [{\citenamefont {Okamoto}(2006)}]{okamoto2006fundamentals}%
  \BibitemOpen
  \bibfield  {author} {\bibinfo {author} {\bibfnamefont {K.}~\bibnamefont
  {Okamoto}},\ }\href@noop {} {\emph {\bibinfo {title} {Fundamentals of optical
  waveguides}}}\ (\bibinfo  {publisher} {Academic press},\ \bibinfo {year}
  {2006})\BibitemShut {NoStop}%
\bibitem [{\citenamefont {Wu}\ \emph {et~al.}(2008)\citenamefont {Wu},
  \citenamefont {Gea-Banacloche},\ and\ \citenamefont
  {Xiao}}]{PhysRevLett.100.173602}%
  \BibitemOpen
  \bibfield  {author} {\bibinfo {author} {\bibfnamefont {H.}~\bibnamefont
  {Wu}}, \bibinfo {author} {\bibfnamefont {J.}~\bibnamefont {Gea-Banacloche}},
  \ and\ \bibinfo {author} {\bibfnamefont {M.}~\bibnamefont {Xiao}},\ }\href
  {\doibase 10.1103/PhysRevLett.100.173602} {\bibfield  {journal} {\bibinfo
  {journal} {Phys. Rev. Lett.}\ }\textbf {\bibinfo {volume} {100}},\ \bibinfo
  {pages} {173602} (\bibinfo {year} {2008})}\BibitemShut {NoStop}%
\bibitem [{\citenamefont {Wang}\ \emph {et~al.}(2001)\citenamefont {Wang},
  \citenamefont {Goorskey},\ and\ \citenamefont
  {Xiao}}]{PhysRevLett.87.073601}%
  \BibitemOpen
  \bibfield  {author} {\bibinfo {author} {\bibfnamefont {H.}~\bibnamefont
  {Wang}}, \bibinfo {author} {\bibfnamefont {D.}~\bibnamefont {Goorskey}}, \
  and\ \bibinfo {author} {\bibfnamefont {M.}~\bibnamefont {Xiao}},\ }\href
  {\doibase 10.1103/PhysRevLett.87.073601} {\bibfield  {journal} {\bibinfo
  {journal} {Phys. Rev. Lett.}\ }\textbf {\bibinfo {volume} {87}},\ \bibinfo
  {pages} {073601} (\bibinfo {year} {2001})}\BibitemShut {NoStop}%
\bibitem [{\citenamefont {Li}\ \emph {et~al.}(2008)\citenamefont {Li},
  \citenamefont {Yang}, \citenamefont {Cao}, \citenamefont {Zhang},
  \citenamefont {Xie},\ and\ \citenamefont {Wang}}]{PhysRevLett.101.073602}%
  \BibitemOpen
  \bibfield  {author} {\bibinfo {author} {\bibfnamefont {S.}~\bibnamefont
  {Li}}, \bibinfo {author} {\bibfnamefont {X.}~\bibnamefont {Yang}}, \bibinfo
  {author} {\bibfnamefont {X.}~\bibnamefont {Cao}}, \bibinfo {author}
  {\bibfnamefont {C.}~\bibnamefont {Zhang}}, \bibinfo {author} {\bibfnamefont
  {C.}~\bibnamefont {Xie}}, \ and\ \bibinfo {author} {\bibfnamefont
  {H.}~\bibnamefont {Wang}},\ }\href {\doibase 10.1103/PhysRevLett.101.073602}
  {\bibfield  {journal} {\bibinfo  {journal} {Phys. Rev. Lett.}\ }\textbf
  {\bibinfo {volume} {101}},\ \bibinfo {pages} {073602} (\bibinfo {year}
  {2008})}\BibitemShut {NoStop}%
\bibitem [{\citenamefont {Wang}\ \emph {et~al.}(2013)\citenamefont {Wang},
  \citenamefont {Wan}, \citenamefont {Zou}, \citenamefont {Zhang},\ and\
  \citenamefont {Zhu}}]{Wang_2013}%
  \BibitemOpen
  \bibfield  {author} {\bibinfo {author} {\bibfnamefont {Y.}~\bibnamefont
  {Wang}}, \bibinfo {author} {\bibfnamefont {J.}~\bibnamefont {Wan}}, \bibinfo
  {author} {\bibfnamefont {B.}~\bibnamefont {Zou}}, \bibinfo {author}
  {\bibfnamefont {J.}~\bibnamefont {Zhang}}, \ and\ \bibinfo {author}
  {\bibfnamefont {Y.}~\bibnamefont {Zhu}},\ }\href {\doibase
  10.1088/1742-6596/414/1/012001} {\bibfield  {journal} {\bibinfo  {journal}
  {Journal of Physics: Conference Series}\ }\textbf {\bibinfo {volume} {414}},\
  \bibinfo {pages} {012001} (\bibinfo {year} {2013})}\BibitemShut {NoStop}%
\bibitem [{\citenamefont {Li}\ \emph {et~al.}(2020{\natexlab{b}})\citenamefont
  {Li}, \citenamefont {Ding}, \citenamefont {Yu}, \citenamefont {Dong},
  \citenamefont {Zeng}, \citenamefont {Zhang}, \citenamefont {Ye},
  \citenamefont {Wu}, \citenamefont {Zhu}, \citenamefont {Gao}, \citenamefont
  {Guo},\ and\ \citenamefont {Shi}}]{PhysRevResearch.2.033517}%
  \BibitemOpen
  \bibfield  {author} {\bibinfo {author} {\bibfnamefont {E.-Z.}\ \bibnamefont
  {Li}}, \bibinfo {author} {\bibfnamefont {D.-S.}\ \bibnamefont {Ding}},
  \bibinfo {author} {\bibfnamefont {Y.-C.}\ \bibnamefont {Yu}}, \bibinfo
  {author} {\bibfnamefont {M.-X.}\ \bibnamefont {Dong}}, \bibinfo {author}
  {\bibfnamefont {L.}~\bibnamefont {Zeng}}, \bibinfo {author} {\bibfnamefont
  {W.-H.}\ \bibnamefont {Zhang}}, \bibinfo {author} {\bibfnamefont {Y.-H.}\
  \bibnamefont {Ye}}, \bibinfo {author} {\bibfnamefont {H.-Z.}\ \bibnamefont
  {Wu}}, \bibinfo {author} {\bibfnamefont {Z.-H.}\ \bibnamefont {Zhu}},
  \bibinfo {author} {\bibfnamefont {W.}~\bibnamefont {Gao}}, \bibinfo {author}
  {\bibfnamefont {G.-C.}\ \bibnamefont {Guo}}, \ and\ \bibinfo {author}
  {\bibfnamefont {B.-S.}\ \bibnamefont {Shi}},\ }\href {\doibase
  10.1103/PhysRevResearch.2.033517} {\bibfield  {journal} {\bibinfo  {journal}
  {Phys. Rev. Res.}\ }\textbf {\bibinfo {volume} {2}},\ \bibinfo {pages}
  {033517} (\bibinfo {year} {2020}{\natexlab{b}})}\BibitemShut {NoStop}%
\bibitem [{\citenamefont {Plenio}\ and\ \citenamefont
  {Knight}(1998)}]{RevModPhys.70.101}%
  \BibitemOpen
  \bibfield  {author} {\bibinfo {author} {\bibfnamefont {M.~B.}\ \bibnamefont
  {Plenio}}\ and\ \bibinfo {author} {\bibfnamefont {P.~L.}\ \bibnamefont
  {Knight}},\ }\href {\doibase 10.1103/RevModPhys.70.101} {\bibfield  {journal}
  {\bibinfo  {journal} {Rev. Mod. Phys.}\ }\textbf {\bibinfo {volume} {70}},\
  \bibinfo {pages} {101} (\bibinfo {year} {1998})}\BibitemShut {NoStop}%
\bibitem [{\citenamefont {Gleyzes}\ \emph {et~al.}(2007)\citenamefont
  {Gleyzes}, \citenamefont {Kuhr}, \citenamefont {Guerlin}, \citenamefont
  {Bernu}, \citenamefont {Del{\'{e}}glise}, \citenamefont {Busk~Hoff},
  \citenamefont {Brune}, \citenamefont {Raimond},\ and\ \citenamefont
  {Haroche}}]{nature05589}%
  \BibitemOpen
  \bibfield  {author} {\bibinfo {author} {\bibfnamefont {S.}~\bibnamefont
  {Gleyzes}}, \bibinfo {author} {\bibfnamefont {S.}~\bibnamefont {Kuhr}},
  \bibinfo {author} {\bibfnamefont {C.}~\bibnamefont {Guerlin}}, \bibinfo
  {author} {\bibfnamefont {J.}~\bibnamefont {Bernu}}, \bibinfo {author}
  {\bibfnamefont {S.}~\bibnamefont {Del{\'{e}}glise}}, \bibinfo {author}
  {\bibfnamefont {U.}~\bibnamefont {Busk~Hoff}}, \bibinfo {author}
  {\bibfnamefont {M.}~\bibnamefont {Brune}}, \bibinfo {author} {\bibfnamefont
  {J.-M.}\ \bibnamefont {Raimond}}, \ and\ \bibinfo {author} {\bibfnamefont
  {S.}~\bibnamefont {Haroche}},\ }\href {\doibase 10.1038/nature05589}
  {\bibfield  {journal} {\bibinfo  {journal} {Nature (London)}\ }\textbf
  {\bibinfo {volume} {446}},\ \bibinfo {pages} {297} (\bibinfo {year}
  {2007})}\BibitemShut {NoStop}%
\bibitem [{\citenamefont {Johansson}\ \emph {et~al.}(2012)\citenamefont
  {Johansson}, \citenamefont {Nation},\ and\ \citenamefont
  {Nori}}]{Com.Phys.Coms.183}%
  \BibitemOpen
  \bibfield  {author} {\bibinfo {author} {\bibfnamefont {J.~R.}\ \bibnamefont
  {Johansson}}, \bibinfo {author} {\bibfnamefont {P.~D.}\ \bibnamefont
  {Nation}}, \ and\ \bibinfo {author} {\bibfnamefont {F.}~\bibnamefont
  {Nori}},\ }\href {\doibase https://doi.org/10.1016/j.cpc.2012.02.021}
  {\bibfield  {journal} {\bibinfo  {journal} {Comput. Phys. Commun.}\ }\textbf
  {\bibinfo {volume} {183}},\ \bibinfo {pages} {1760} (\bibinfo {year}
  {2012})}\BibitemShut {NoStop}%
\bibitem [{\citenamefont {Morin}\ \emph {et~al.}(2019)\citenamefont {Morin},
  \citenamefont {K\"orber}, \citenamefont {Langenfeld},\ and\ \citenamefont
  {Rempe}}]{PhysRevLett.123.133602}%
  \BibitemOpen
  \bibfield  {author} {\bibinfo {author} {\bibfnamefont {O.}~\bibnamefont
  {Morin}}, \bibinfo {author} {\bibfnamefont {M.}~\bibnamefont {K\"orber}},
  \bibinfo {author} {\bibfnamefont {S.}~\bibnamefont {Langenfeld}}, \ and\
  \bibinfo {author} {\bibfnamefont {G.}~\bibnamefont {Rempe}},\ }\href
  {\doibase 10.1103/PhysRevLett.123.133602} {\bibfield  {journal} {\bibinfo
  {journal} {Phys. Rev. Lett.}\ }\textbf {\bibinfo {volume} {123}},\ \bibinfo
  {pages} {133602} (\bibinfo {year} {2019})}\BibitemShut {NoStop}%
\bibitem [{\citenamefont {Luo}\ \emph {et~al.}(2015)\citenamefont {Luo},
  \citenamefont {Herrmann}, \citenamefont {Krapick}, \citenamefont {Brecht},
  \citenamefont {Ricken}, \citenamefont {Quiring}, \citenamefont {Suche},
  \citenamefont {Sohler},\ and\ \citenamefont {Silberhorn}}]{Luo_2015}%
  \BibitemOpen
  \bibfield  {author} {\bibinfo {author} {\bibfnamefont {K.-H.}\ \bibnamefont
  {Luo}}, \bibinfo {author} {\bibfnamefont {H.}~\bibnamefont {Herrmann}},
  \bibinfo {author} {\bibfnamefont {S.}~\bibnamefont {Krapick}}, \bibinfo
  {author} {\bibfnamefont {B.}~\bibnamefont {Brecht}}, \bibinfo {author}
  {\bibfnamefont {R.}~\bibnamefont {Ricken}}, \bibinfo {author} {\bibfnamefont
  {V.}~\bibnamefont {Quiring}}, \bibinfo {author} {\bibfnamefont
  {H.}~\bibnamefont {Suche}}, \bibinfo {author} {\bibfnamefont
  {W.}~\bibnamefont {Sohler}}, \ and\ \bibinfo {author} {\bibfnamefont
  {C.}~\bibnamefont {Silberhorn}},\ }\href {\doibase
  10.1088/1367-2630/17/7/073039} {\bibfield  {journal} {\bibinfo  {journal}
  {New J. Phys.}\ }\textbf {\bibinfo {volume} {17}},\ \bibinfo {pages} {073039}
  (\bibinfo {year} {2015})}\BibitemShut {NoStop}%
\bibitem [{\citenamefont {Wolfgramm}\ \emph {et~al.}(2008)\citenamefont
  {Wolfgramm}, \citenamefont {Xing}, \citenamefont {Cer\`{e}}, \citenamefont
  {Predojevi\'{c}}, \citenamefont {Steinberg},\ and\ \citenamefont
  {Mitchell}}]{OE.16.018145}%
  \BibitemOpen
  \bibfield  {author} {\bibinfo {author} {\bibfnamefont {F.}~\bibnamefont
  {Wolfgramm}}, \bibinfo {author} {\bibfnamefont {X.}~\bibnamefont {Xing}},
  \bibinfo {author} {\bibfnamefont {A.}~\bibnamefont {Cer\`{e}}}, \bibinfo
  {author} {\bibfnamefont {A.}~\bibnamefont {Predojevi\'{c}}}, \bibinfo
  {author} {\bibfnamefont {A.~M.}\ \bibnamefont {Steinberg}}, \ and\ \bibinfo
  {author} {\bibfnamefont {M.~W.}\ \bibnamefont {Mitchell}},\ }\href {\doibase
  10.1364/OE.16.018145} {\bibfield  {journal} {\bibinfo  {journal} {Opt.
  Express}\ }\textbf {\bibinfo {volume} {16}},\ \bibinfo {pages} {18145}
  (\bibinfo {year} {2008})}\BibitemShut {NoStop}%
\bibitem [{\citenamefont {Le~Jeannic}\ \emph {et~al.}(2016)\citenamefont
  {Le~Jeannic}, \citenamefont {Verma}, \citenamefont {Cavailles}, \citenamefont
  {Marsili}, \citenamefont {Shaw}, \citenamefont {Huang}, \citenamefont
  {Morin}, \citenamefont {Nam},\ and\ \citenamefont {Laurat}}]{OL.41.005341}%
  \BibitemOpen
  \bibfield  {author} {\bibinfo {author} {\bibfnamefont {H.}~\bibnamefont
  {Le~Jeannic}}, \bibinfo {author} {\bibfnamefont {V.~B.}\ \bibnamefont
  {Verma}}, \bibinfo {author} {\bibfnamefont {A.}~\bibnamefont {Cavailles}},
  \bibinfo {author} {\bibfnamefont {F.}~\bibnamefont {Marsili}}, \bibinfo
  {author} {\bibfnamefont {M.~D.}\ \bibnamefont {Shaw}}, \bibinfo {author}
  {\bibfnamefont {K.}~\bibnamefont {Huang}}, \bibinfo {author} {\bibfnamefont
  {O.}~\bibnamefont {Morin}}, \bibinfo {author} {\bibfnamefont {S.~W.}\
  \bibnamefont {Nam}}, \ and\ \bibinfo {author} {\bibfnamefont
  {J.}~\bibnamefont {Laurat}},\ }\href {\doibase 10.1364/OL.41.005341}
  {\bibfield  {journal} {\bibinfo  {journal} {Opt. Lett.}\ }\textbf {\bibinfo
  {volume} {41}},\ \bibinfo {pages} {5341} (\bibinfo {year}
  {2016})}\BibitemShut {NoStop}%
\bibitem [{\citenamefont {Fasel}\ \emph {et~al.}(2004)\citenamefont {Fasel},
  \citenamefont {Alibart}, \citenamefont {Tanzilli}, \citenamefont {Baldi},
  \citenamefont {Beveratos}, \citenamefont {Gisin},\ and\ \citenamefont
  {Zbinden}}]{Fasel_2004}%
  \BibitemOpen
  \bibfield  {author} {\bibinfo {author} {\bibfnamefont {S.}~\bibnamefont
  {Fasel}}, \bibinfo {author} {\bibfnamefont {O.}~\bibnamefont {Alibart}},
  \bibinfo {author} {\bibfnamefont {S.}~\bibnamefont {Tanzilli}}, \bibinfo
  {author} {\bibfnamefont {P.}~\bibnamefont {Baldi}}, \bibinfo {author}
  {\bibfnamefont {A.}~\bibnamefont {Beveratos}}, \bibinfo {author}
  {\bibfnamefont {N.}~\bibnamefont {Gisin}}, \ and\ \bibinfo {author}
  {\bibfnamefont {H.}~\bibnamefont {Zbinden}},\ }\href {\doibase
  10.1088/1367-2630/6/1/163} {\bibfield  {journal} {\bibinfo  {journal} {New J.
  Phys.}\ }\textbf {\bibinfo {volume} {6}},\ \bibinfo {pages} {163} (\bibinfo
  {year} {2004})}\BibitemShut {NoStop}%
\bibitem [{\citenamefont {Scholz}\ \emph {et~al.}(2009)\citenamefont {Scholz},
  \citenamefont {Koch},\ and\ \citenamefont {Benson}}]{PhysRevLett.102.063603}%
  \BibitemOpen
  \bibfield  {author} {\bibinfo {author} {\bibfnamefont {M.}~\bibnamefont
  {Scholz}}, \bibinfo {author} {\bibfnamefont {L.}~\bibnamefont {Koch}}, \ and\
  \bibinfo {author} {\bibfnamefont {O.}~\bibnamefont {Benson}},\ }\href
  {\doibase 10.1103/PhysRevLett.102.063603} {\bibfield  {journal} {\bibinfo
  {journal} {Phys. Rev. Lett.}\ }\textbf {\bibinfo {volume} {102}},\ \bibinfo
  {pages} {063603} (\bibinfo {year} {2009})}\BibitemShut {NoStop}%
\bibitem [{\citenamefont {Bocquillon}\ \emph {et~al.}(2009)\citenamefont
  {Bocquillon}, \citenamefont {Couteau}, \citenamefont {Razavi}, \citenamefont
  {Laflamme},\ and\ \citenamefont {Weihs}}]{PhysRevA.79.035801}%
  \BibitemOpen
  \bibfield  {author} {\bibinfo {author} {\bibfnamefont {E.}~\bibnamefont
  {Bocquillon}}, \bibinfo {author} {\bibfnamefont {C.}~\bibnamefont {Couteau}},
  \bibinfo {author} {\bibfnamefont {M.}~\bibnamefont {Razavi}}, \bibinfo
  {author} {\bibfnamefont {R.}~\bibnamefont {Laflamme}}, \ and\ \bibinfo
  {author} {\bibfnamefont {G.}~\bibnamefont {Weihs}},\ }\href {\doibase
  10.1103/PhysRevA.79.035801} {\bibfield  {journal} {\bibinfo  {journal} {Phys.
  Rev. A}\ }\textbf {\bibinfo {volume} {79}},\ \bibinfo {pages} {035801}
  (\bibinfo {year} {2009})}\BibitemShut {NoStop}%
\bibitem [{\citenamefont {Li}\ \emph {et~al.}(2020{\natexlab{c}})\citenamefont
  {Li}, \citenamefont {Su}, \citenamefont {Cui}, \citenamefont {Xie},
  \citenamefont {Ou},\ and\ \citenamefont {Li}}]{APL.5.0003601}%
  \BibitemOpen
  \bibfield  {author} {\bibinfo {author} {\bibfnamefont {J.}~\bibnamefont
  {Li}}, \bibinfo {author} {\bibfnamefont {J.}~\bibnamefont {Su}}, \bibinfo
  {author} {\bibfnamefont {L.}~\bibnamefont {Cui}}, \bibinfo {author}
  {\bibfnamefont {T.}~\bibnamefont {Xie}}, \bibinfo {author} {\bibfnamefont
  {Z.~Y.}\ \bibnamefont {Ou}}, \ and\ \bibinfo {author} {\bibfnamefont
  {X.}~\bibnamefont {Li}},\ }\href@noop {} {\bibfield  {journal} {\bibinfo
  {journal} {Appl. Phys. Lett.}\ }\textbf {\bibinfo {volume} {116}} (\bibinfo
  {year} {2020}{\natexlab{c}})}\BibitemShut {NoStop}%
\bibitem [{\citenamefont {Slussarenko}\ \emph {et~al.}(2017)\citenamefont
  {Slussarenko}, \citenamefont {Weston}, \citenamefont {Chrzanowski},
  \citenamefont {Shalm}, \citenamefont {Verma}, \citenamefont {Nam},\ and\
  \citenamefont {Pryde}}]{nphoton.2017.10.1038}%
  \BibitemOpen
  \bibfield  {author} {\bibinfo {author} {\bibfnamefont {S.}~\bibnamefont
  {Slussarenko}}, \bibinfo {author} {\bibfnamefont {M.~M.}\ \bibnamefont
  {Weston}}, \bibinfo {author} {\bibfnamefont {H.~M.}\ \bibnamefont
  {Chrzanowski}}, \bibinfo {author} {\bibfnamefont {L.~K.}\ \bibnamefont
  {Shalm}}, \bibinfo {author} {\bibfnamefont {V.~B.}\ \bibnamefont {Verma}},
  \bibinfo {author} {\bibfnamefont {S.~W.}\ \bibnamefont {Nam}}, \ and\
  \bibinfo {author} {\bibfnamefont {G.~J.}\ \bibnamefont {Pryde}},\ }\href
  {\doibase 10.1038/s41566-017-0011-5} {\bibfield  {journal} {\bibinfo
  {journal} {Nat. Photonics}\ }\textbf {\bibinfo {volume} {11}},\ \bibinfo
  {pages} {700} (\bibinfo {year} {2017})}\BibitemShut {NoStop}%
\bibitem [{\citenamefont {Broome}\ \emph {et~al.}(2011)\citenamefont {Broome},
  \citenamefont {Almeida}, \citenamefont {Fedrizzi},\ and\ \citenamefont
  {White}}]{OE.19.022698}%
  \BibitemOpen
  \bibfield  {author} {\bibinfo {author} {\bibfnamefont {M.~A.}\ \bibnamefont
  {Broome}}, \bibinfo {author} {\bibfnamefont {M.~P.}\ \bibnamefont {Almeida}},
  \bibinfo {author} {\bibfnamefont {A.}~\bibnamefont {Fedrizzi}}, \ and\
  \bibinfo {author} {\bibfnamefont {A.~G.}\ \bibnamefont {White}},\ }\href
  {\doibase 10.1364/OE.19.022698} {\bibfield  {journal} {\bibinfo  {journal}
  {Opt. Express}\ }\textbf {\bibinfo {volume} {19}},\ \bibinfo {pages} {22698}
  (\bibinfo {year} {2011})}\BibitemShut {NoStop}%
\bibitem [{\citenamefont {Tomm}\ \emph {et~al.}(2021)\citenamefont {Tomm},
  \citenamefont {Javadi}, \citenamefont {Antoniadis}, \citenamefont {Najer},
  \citenamefont {L\"{o}bl}, \citenamefont {Korsch}, \citenamefont {Schott},
  \citenamefont {Valentin}, \citenamefont {Wieck}, \citenamefont {Ludwig},\
  and\ \citenamefont {Warburton}}]{nnanotechnol.10.1038.108}%
  \BibitemOpen
  \bibfield  {author} {\bibinfo {author} {\bibfnamefont {N.}~\bibnamefont
  {Tomm}}, \bibinfo {author} {\bibfnamefont {A.}~\bibnamefont {Javadi}},
  \bibinfo {author} {\bibfnamefont {N.~O.}\ \bibnamefont {Antoniadis}},
  \bibinfo {author} {\bibfnamefont {D.}~\bibnamefont {Najer}}, \bibinfo
  {author} {\bibfnamefont {M.~C.}\ \bibnamefont {L\"{o}bl}}, \bibinfo {author}
  {\bibfnamefont {A.~R.}\ \bibnamefont {Korsch}}, \bibinfo {author}
  {\bibfnamefont {R.}~\bibnamefont {Schott}}, \bibinfo {author} {\bibfnamefont
  {S.~R.}\ \bibnamefont {Valentin}}, \bibinfo {author} {\bibfnamefont {A.~D.}\
  \bibnamefont {Wieck}}, \bibinfo {author} {\bibfnamefont {A.}~\bibnamefont
  {Ludwig}}, \ and\ \bibinfo {author} {\bibfnamefont {R.~J.}\ \bibnamefont
  {Warburton}},\ }\href {https://doi.org/10.1038/s41565-020-00831-x} {\bibfield
   {journal} {\bibinfo  {journal} {Nat. Nanotechnol.}\ } (\bibinfo {year}
  {2021})}\BibitemShut {NoStop}%
\bibitem [{\citenamefont {Eisaman}\ \emph {et~al.}(2011)\citenamefont
  {Eisaman}, \citenamefont {Fan}, \citenamefont {Migdall},\ and\ \citenamefont
  {Polyakov}}]{Rew.Sci.Instrum.82.071101}%
  \BibitemOpen
  \bibfield  {author} {\bibinfo {author} {\bibfnamefont {M.~D.}\ \bibnamefont
  {Eisaman}}, \bibinfo {author} {\bibfnamefont {J.}~\bibnamefont {Fan}},
  \bibinfo {author} {\bibfnamefont {A.}~\bibnamefont {Migdall}}, \ and\
  \bibinfo {author} {\bibfnamefont {S.~V.}\ \bibnamefont {Polyakov}},\ }\href
  {\doibase 10.1063/1.3610677} {\bibfield  {journal} {\bibinfo  {journal} {Rev.
  Sci. Instrum.}\ }\textbf {\bibinfo {volume} {82}},\ \bibinfo {pages} {071101}
  (\bibinfo {year} {2011})}\BibitemShut {NoStop}%
\bibitem [{\citenamefont {Fischer}\ \emph {et~al.}(2017)\citenamefont
  {Fischer}, \citenamefont {Hanschke}, \citenamefont {Wierzbowski},
  \citenamefont {Simmet}, \citenamefont {Dory}, \citenamefont {Finley},
  \citenamefont {Vu\v{c}kovi\'c},\ and\ \citenamefont
  {M\"{u}~ller}}]{nphys4052}%
  \BibitemOpen
  \bibfield  {author} {\bibinfo {author} {\bibfnamefont {K.}~\bibnamefont
  {Fischer}}, \bibinfo {author} {\bibfnamefont {L.}~\bibnamefont {Hanschke}},
  \bibinfo {author} {\bibfnamefont {J.}~\bibnamefont {Wierzbowski}}, \bibinfo
  {author} {\bibfnamefont {T.}~\bibnamefont {Simmet}}, \bibinfo {author}
  {\bibfnamefont {C.}~\bibnamefont {Dory}}, \bibinfo {author} {\bibfnamefont
  {J.}~\bibnamefont {Finley}}, \bibinfo {author} {\bibfnamefont
  {J.}~\bibnamefont {Vu\v{c}kovi\'c}}, \ and\ \bibinfo {author} {\bibfnamefont
  {K.}~\bibnamefont {M\"{u}~ller}},\ }\href {\doibase 10.1038/nphys4052}
  {\bibfield  {journal} {\bibinfo  {journal} {Nat. Phys.}\ }\textbf {\bibinfo
  {volume} {13}},\ \bibinfo {pages} {649} (\bibinfo {year} {2017})}\BibitemShut
  {NoStop}%
\bibitem [{\citenamefont {{n}oz}\ \emph {et~al.}(2018)\citenamefont {{n}oz},
  \citenamefont {Laussy}, \citenamefont {del Valle}, \citenamefont {Tejedor},\
  and\ \citenamefont {Gonz\'{a}lez-Tudela}}]{SanchezMunoz:18}%
  \BibitemOpen
  \bibfield  {author} {\bibinfo {author} {\bibfnamefont {C.~S.~M.}\
  \bibnamefont {{n}oz}}, \bibinfo {author} {\bibfnamefont {F.~P.}\ \bibnamefont
  {Laussy}}, \bibinfo {author} {\bibfnamefont {E.}~\bibnamefont {del Valle}},
  \bibinfo {author} {\bibfnamefont {C.}~\bibnamefont {Tejedor}}, \ and\
  \bibinfo {author} {\bibfnamefont {A.}~\bibnamefont {Gonz\'{a}lez-Tudela}},\
  }\href {\doibase 10.1364/OPTICA.5.000014} {\bibfield  {journal} {\bibinfo
  {journal} {Optica}\ }\textbf {\bibinfo {volume} {5}},\ \bibinfo {pages} {14}
  (\bibinfo {year} {2018})}\BibitemShut {NoStop}%
\bibitem [{\citenamefont {Schweickert}\ \emph {et~al.}(2018)\citenamefont
  {Schweickert}, \citenamefont {J\"ons}, \citenamefont {Zeuner}, \citenamefont
  {Covre~da Silva}, \citenamefont {Huang}, \citenamefont {Lettner},
  \citenamefont {Reindl}, \citenamefont {Zichi}, \citenamefont {Trotta},
  \citenamefont {Rastelli},\ and\ \citenamefont {Zwiller}}]{APL.1.5020038}%
  \BibitemOpen
  \bibfield  {author} {\bibinfo {author} {\bibfnamefont {L.}~\bibnamefont
  {Schweickert}}, \bibinfo {author} {\bibfnamefont {K.~D.}\ \bibnamefont
  {J\"ons}}, \bibinfo {author} {\bibfnamefont {K.~D.}\ \bibnamefont {Zeuner}},
  \bibinfo {author} {\bibfnamefont {S.~F.}\ \bibnamefont {Covre~da Silva}},
  \bibinfo {author} {\bibfnamefont {H.}~\bibnamefont {Huang}}, \bibinfo
  {author} {\bibfnamefont {T.}~\bibnamefont {Lettner}}, \bibinfo {author}
  {\bibfnamefont {M.}~\bibnamefont {Reindl}}, \bibinfo {author} {\bibfnamefont
  {J.}~\bibnamefont {Zichi}}, \bibinfo {author} {\bibfnamefont
  {R.}~\bibnamefont {Trotta}}, \bibinfo {author} {\bibfnamefont
  {A.}~\bibnamefont {Rastelli}}, \ and\ \bibinfo {author} {\bibfnamefont
  {V.}~\bibnamefont {Zwiller}},\ }\href@noop {} {\bibfield  {journal} {\bibinfo
   {journal} {Appl. Phys. Lett.}\ }\textbf {\bibinfo {volume} {112}},\ \bibinfo
  {pages} {093106} (\bibinfo {year} {2018})}\BibitemShut {NoStop}%
\bibitem [{\citenamefont {Yan}\ \emph {et~al.}(2019)\citenamefont {Yan},
  \citenamefont {Zhang}, \citenamefont {Gong}, \citenamefont {Wu},
  \citenamefont {Zheng}, \citenamefont {Li}, \citenamefont {Wang},
  \citenamefont {Liang}, \citenamefont {Lin}, \citenamefont {Xu}, \citenamefont
  {Guo}, \citenamefont {Sun}, \citenamefont {Peng}, \citenamefont {Xia},
  \citenamefont {Deng}, \citenamefont {Rong}, \citenamefont {You},
  \citenamefont {Nori}, \citenamefont {Fan}, \citenamefont {Zhu},\ and\
  \citenamefont {Pan}}]{science.aaw1611}%
  \BibitemOpen
  \bibfield  {author} {\bibinfo {author} {\bibfnamefont {Z.}~\bibnamefont
  {Yan}}, \bibinfo {author} {\bibfnamefont {Y.-R.}\ \bibnamefont {Zhang}},
  \bibinfo {author} {\bibfnamefont {M.}~\bibnamefont {Gong}}, \bibinfo {author}
  {\bibfnamefont {Y.}~\bibnamefont {Wu}}, \bibinfo {author} {\bibfnamefont
  {Y.}~\bibnamefont {Zheng}}, \bibinfo {author} {\bibfnamefont
  {S.}~\bibnamefont {Li}}, \bibinfo {author} {\bibfnamefont {C.}~\bibnamefont
  {Wang}}, \bibinfo {author} {\bibfnamefont {F.}~\bibnamefont {Liang}},
  \bibinfo {author} {\bibfnamefont {J.}~\bibnamefont {Lin}}, \bibinfo {author}
  {\bibfnamefont {Y.}~\bibnamefont {Xu}}, \bibinfo {author} {\bibfnamefont
  {C.}~\bibnamefont {Guo}}, \bibinfo {author} {\bibfnamefont {L.}~\bibnamefont
  {Sun}}, \bibinfo {author} {\bibfnamefont {C.-Z.}\ \bibnamefont {Peng}},
  \bibinfo {author} {\bibfnamefont {K.}~\bibnamefont {Xia}}, \bibinfo {author}
  {\bibfnamefont {H.}~\bibnamefont {Deng}}, \bibinfo {author} {\bibfnamefont
  {H.}~\bibnamefont {Rong}}, \bibinfo {author} {\bibfnamefont {J.~Q.}\
  \bibnamefont {You}}, \bibinfo {author} {\bibfnamefont {F.}~\bibnamefont
  {Nori}}, \bibinfo {author} {\bibfnamefont {H.}~\bibnamefont {Fan}}, \bibinfo
  {author} {\bibfnamefont {X.}~\bibnamefont {Zhu}}, \ and\ \bibinfo {author}
  {\bibfnamefont {J.-W.}\ \bibnamefont {Pan}},\ }\href {\doibase
  10.1126/science.aaw1611} {\bibfield  {journal} {\bibinfo  {journal}
  {Science}\ }\textbf {\bibinfo {volume} {364}},\ \bibinfo {pages} {753}
  (\bibinfo {year} {2019})}\BibitemShut {NoStop}%
\bibitem [{\citenamefont {Xiao}\ \emph {et~al.}(1995)\citenamefont {Xiao},
  \citenamefont {Li}, \citenamefont {Jin},\ and\ \citenamefont
  {Gea-Banacloche}}]{PhysRevLett.74.666}%
  \BibitemOpen
  \bibfield  {author} {\bibinfo {author} {\bibfnamefont {M.}~\bibnamefont
  {Xiao}}, \bibinfo {author} {\bibfnamefont {Y.-q.}\ \bibnamefont {Li}},
  \bibinfo {author} {\bibfnamefont {S.-z.}\ \bibnamefont {Jin}}, \ and\
  \bibinfo {author} {\bibfnamefont {J.}~\bibnamefont {Gea-Banacloche}},\ }\href
  {\doibase 10.1103/PhysRevLett.74.666} {\bibfield  {journal} {\bibinfo
  {journal} {Phys. Rev. Lett.}\ }\textbf {\bibinfo {volume} {74}},\ \bibinfo
  {pages} {666} (\bibinfo {year} {1995})}\BibitemShut {NoStop}%
\bibitem [{\citenamefont {Lilienfein}\ \emph {et~al.}(2017)\citenamefont
  {Lilienfein}, \citenamefont {Hofer}, \citenamefont {Holzberger},
  \citenamefont {Matzer}, \citenamefont {Zimmermann}, \citenamefont
  {Trubetskov}, \citenamefont {Pervak},\ and\ \citenamefont
  {Pupeza}}]{OL.42.000271}%
  \BibitemOpen
  \bibfield  {author} {\bibinfo {author} {\bibfnamefont {N.}~\bibnamefont
  {Lilienfein}}, \bibinfo {author} {\bibfnamefont {C.}~\bibnamefont {Hofer}},
  \bibinfo {author} {\bibfnamefont {S.}~\bibnamefont {Holzberger}}, \bibinfo
  {author} {\bibfnamefont {C.}~\bibnamefont {Matzer}}, \bibinfo {author}
  {\bibfnamefont {P.}~\bibnamefont {Zimmermann}}, \bibinfo {author}
  {\bibfnamefont {M.}~\bibnamefont {Trubetskov}}, \bibinfo {author}
  {\bibfnamefont {V.}~\bibnamefont {Pervak}}, \ and\ \bibinfo {author}
  {\bibfnamefont {I.}~\bibnamefont {Pupeza}},\ }\href {\doibase
  10.1364/OL.42.000271} {\bibfield  {journal} {\bibinfo  {journal} {Opt.
  Lett.}\ }\textbf {\bibinfo {volume} {42}},\ \bibinfo {pages} {271} (\bibinfo
  {year} {2017})}\BibitemShut {NoStop}%
\bibitem [{\citenamefont {Zhang}\ \emph {et~al.}(2018)\citenamefont {Zhang},
  \citenamefont {Hu}, \citenamefont {Lin}, \citenamefont {Niu}, \citenamefont
  {Xia}, \citenamefont {Gong},\ and\ \citenamefont {Gong}}]{nphoton.2018.1038}%
  \BibitemOpen
  \bibfield  {author} {\bibinfo {author} {\bibfnamefont {S.}~\bibnamefont
  {Zhang}}, \bibinfo {author} {\bibfnamefont {Y.}~\bibnamefont {Hu}}, \bibinfo
  {author} {\bibfnamefont {G.}~\bibnamefont {Lin}}, \bibinfo {author}
  {\bibfnamefont {Y.}~\bibnamefont {Niu}}, \bibinfo {author} {\bibfnamefont
  {K.}~\bibnamefont {Xia}}, \bibinfo {author} {\bibfnamefont {J.}~\bibnamefont
  {Gong}}, \ and\ \bibinfo {author} {\bibfnamefont {S.}~\bibnamefont {Gong}},\
  }\href {\doibase 10.1038/s41566-018-0269-2} {\bibfield  {journal} {\bibinfo
  {journal} {Nat. Photonics}\ }\textbf {\bibinfo {volume} {12}},\ \bibinfo
  {pages} {744} (\bibinfo {year} {2018})}\BibitemShut {NoStop}%
\bibitem [{\citenamefont {Muller}\ \emph {et~al.}(2010)\citenamefont {Muller},
  \citenamefont {Flagg}, \citenamefont {Lawall},\ and\ \citenamefont
  {Solomon}}]{OL.35.002293}%
  \BibitemOpen
  \bibfield  {author} {\bibinfo {author} {\bibfnamefont {A.}~\bibnamefont
  {Muller}}, \bibinfo {author} {\bibfnamefont {E.~B.}\ \bibnamefont {Flagg}},
  \bibinfo {author} {\bibfnamefont {J.~R.}\ \bibnamefont {Lawall}}, \ and\
  \bibinfo {author} {\bibfnamefont {G.~S.}\ \bibnamefont {Solomon}},\ }\href
  {\doibase 10.1364/OL.35.002293} {\bibfield  {journal} {\bibinfo  {journal}
  {Opt. Lett.}\ }\textbf {\bibinfo {volume} {35}},\ \bibinfo {pages} {2293}
  (\bibinfo {year} {2010})}\BibitemShut {NoStop}%
\bibitem [{\citenamefont {Zhang}\ \emph {et~al.}(2013)\citenamefont {Zhang},
  \citenamefont {Wang}, \citenamefont {Chen}, \citenamefont {Bao},
  \citenamefont {Zhang}, \citenamefont {Zhao},\ and\ \citenamefont
  {Jia}}]{PhysRevA.87.033835}%
  \BibitemOpen
  \bibfield  {author} {\bibinfo {author} {\bibfnamefont {H.}~\bibnamefont
  {Zhang}}, \bibinfo {author} {\bibfnamefont {L.}~\bibnamefont {Wang}},
  \bibinfo {author} {\bibfnamefont {J.}~\bibnamefont {Chen}}, \bibinfo {author}
  {\bibfnamefont {S.}~\bibnamefont {Bao}}, \bibinfo {author} {\bibfnamefont
  {L.}~\bibnamefont {Zhang}}, \bibinfo {author} {\bibfnamefont
  {J.}~\bibnamefont {Zhao}}, \ and\ \bibinfo {author} {\bibfnamefont
  {S.}~\bibnamefont {Jia}},\ }\href {\doibase 10.1103/PhysRevA.87.033835}
  {\bibfield  {journal} {\bibinfo  {journal} {Phys. Rev. A}\ }\textbf {\bibinfo
  {volume} {87}},\ \bibinfo {pages} {033835} (\bibinfo {year}
  {2013})}\BibitemShut {NoStop}%
\bibitem [{\citenamefont {Stern}\ \emph {et~al.}(2013)\citenamefont {Stern},
  \citenamefont {Desiatov}, \citenamefont {Goykhman},\ and\ \citenamefont
  {Levy}}]{ncomms2554}%
  \BibitemOpen
  \bibfield  {author} {\bibinfo {author} {\bibfnamefont {L.}~\bibnamefont
  {Stern}}, \bibinfo {author} {\bibfnamefont {B.}~\bibnamefont {Desiatov}},
  \bibinfo {author} {\bibfnamefont {I.}~\bibnamefont {Goykhman}}, \ and\
  \bibinfo {author} {\bibfnamefont {U.}~\bibnamefont {Levy}},\ }\href {\doibase
  10.1038/ncomms2554} {\bibfield  {journal} {\bibinfo  {journal} {Nat.
  Commun.}\ }\textbf {\bibinfo {volume} {4}},\ \bibinfo {pages} {1548}
  (\bibinfo {year} {2013})}\BibitemShut {NoStop}%
\bibitem [{\citenamefont {Ritter}\ \emph {et~al.}(2015)\citenamefont {Ritter},
  \citenamefont {Gruhler}, \citenamefont {Pernice}, \citenamefont {K\"ubler},
  \citenamefont {Pfau},\ and\ \citenamefont {L\"ow}}]{APL.1.4927172}%
  \BibitemOpen
  \bibfield  {author} {\bibinfo {author} {\bibfnamefont {R.}~\bibnamefont
  {Ritter}}, \bibinfo {author} {\bibfnamefont {N.}~\bibnamefont {Gruhler}},
  \bibinfo {author} {\bibfnamefont {W.}~\bibnamefont {Pernice}}, \bibinfo
  {author} {\bibfnamefont {H.}~\bibnamefont {K\"ubler}}, \bibinfo {author}
  {\bibfnamefont {T.}~\bibnamefont {Pfau}}, \ and\ \bibinfo {author}
  {\bibfnamefont {R.}~\bibnamefont {L\"ow}},\ }\href {\doibase
  10.1063/1.4927172} {\bibfield  {journal} {\bibinfo  {journal} {Appl. Phys.
  Lett.}\ }\textbf {\bibinfo {volume} {107}},\ \bibinfo {pages} {041101}
  (\bibinfo {year} {2015})}\BibitemShut {NoStop}%
\bibitem [{\citenamefont {Ritter}\ \emph {et~al.}(2018)\citenamefont {Ritter},
  \citenamefont {Gruhler}, \citenamefont {Dobbertin}, \citenamefont {K\"ubler},
  \citenamefont {Scheel}, \citenamefont {Pernice}, \citenamefont {Pfau},\ and\
  \citenamefont {L\"ow}}]{PhysRevX.8.021032}%
  \BibitemOpen
  \bibfield  {author} {\bibinfo {author} {\bibfnamefont {R.}~\bibnamefont
  {Ritter}}, \bibinfo {author} {\bibfnamefont {N.}~\bibnamefont {Gruhler}},
  \bibinfo {author} {\bibfnamefont {H.}~\bibnamefont {Dobbertin}}, \bibinfo
  {author} {\bibfnamefont {H.}~\bibnamefont {K\"ubler}}, \bibinfo {author}
  {\bibfnamefont {S.}~\bibnamefont {Scheel}}, \bibinfo {author} {\bibfnamefont
  {W.}~\bibnamefont {Pernice}}, \bibinfo {author} {\bibfnamefont
  {T.}~\bibnamefont {Pfau}}, \ and\ \bibinfo {author} {\bibfnamefont
  {R.}~\bibnamefont {L\"ow}},\ }\href {\doibase 10.1103/PhysRevX.8.021032}
  {\bibfield  {journal} {\bibinfo  {journal} {Phys. Rev. X}\ }\textbf {\bibinfo
  {volume} {8}},\ \bibinfo {pages} {021032} (\bibinfo {year}
  {2018})}\BibitemShut {NoStop}%
\end{thebibliography}

%

\end{document}